\documentclass[aps, prd, twocolumn, showpacs, superscriptaddress, groupedaddress]{revtex4-1} 
\usepackage{graphicx}	
\usepackage{amssymb}
\usepackage{dcolumn}
\usepackage{color}
\usepackage{subfigure, rotating, bm, array}
\usepackage[pagebackref=false, colorlinks=true]{hyperref}
\hypersetup{
linkcolor=blue,     % color of internal links
citecolor=blue,     % color of links to bibliography
urlcolor=blue} 
\begin{document}

\title{Shadows and precession of orbits in rotating Janis-Newman-Winicour spacetime}
\author{Divyesh N. Solanki}
\email{divyeshsolanki98@gmail.com}
\affiliation{Sardar Vallabhbhai National Institute of Technology, Surat GUJ 395007,  India}
\author{Parth Bambhaniya}
\email{grcollapse@gmail.com}
\affiliation{International Center for Cosmology, Charusat University, Anand, GUJ 388421, India}
\author{Dipanjan Dey}
\email{dipanjandey.icc@charusat.ac.in}
\affiliation{International Center for Cosmology, Charusat University, Anand, GUJ 388421, India}
\author{Pankaj S. Joshi}
\email{psjprovost@charusat.ac.in}
\affiliation{International Center for Cosmology, Charusat University, Anand, GUJ 388421, India}
\author{Kamlesh N. Pathak}
\email{knp@phy.svnit.ac.in}
\affiliation{Department of Physics, S.V National Institute of Technology, Surat GUJ 395007,  India}\date{\today}

\begin{abstract}
In this paper, we construct the rotating Janis-Newman-Winicour (JNW) naked singularity spacetime using Newman-Janis Algorithm (NJA). We analyse NJA with and without complexification methods and find that the energy conditions do satisfied when we skip the complexification step. We study the shadows cast by rotating JNW naked singularity and compare them with the shadows cast by the Kerr black hole. We find that the shadow of the rotating naked singularity can be distinguished from the shadow of the Kerr black hole. While we analyse the precession of timelike bound orbits in rotating JNW spacetime, we find that it can have a negative (or opposite) precession, which is not present in the Kerr black hole case. These novel signatures of the shadow and orbital precession in rotating JNW naked singularity spacetime could be important in the context of the recent observation of the shadow of the M87 galactic center and the stellar dynamics of `S-stars' around Milkyway galactic center.

\bigskip
Key words: Black hole, Naked singularity, Shadow, Timelike orbits, Newman-Janis Algorithm.
\end{abstract}
\maketitle

\section{Introduction}

In 1965, Newman and Janis showed that the Kerr metric could be obtained from the Schwarzschild metric using a complex coordinate transformation \cite{JNA}. For the static Reissner-Nordström metric, using the same method, one can derive the Kerr-Newman metric, which represents a spacetime geometry of the electrically charged and rotating black hole \cite{KN}. The application of the NJA for generating interior solutions which match smoothly to the external Kerr metric is studied in \cite{Drake1997}. In this algorithm, authors use two different types of complexification with no apparent reason. There are many ways to complexify the coordinates, however, in the above mentioned literature, the only reason given for choosing those particular complexifications is that it is successful in generating the Kerr and Kerr-Newman metric. However, the success of the NJA is limited to the spacetimes which atleast satisfy the reciprocal condition, $g_{tt}g_{rr} = -1$. Later, it was proved by S.P. Drake, and P. Szekeres \cite{Drake2000} that the only perfect fluid solution generated by the NJA is the Kerr metric, and the only Petrov typed D solution to the Einstein-Maxwell equation is the Kerr-Newman metric.

The general theory of relativity predicts that when large enough masses collapse under the influence of their own gravity, the spacetime singularity forms necessarily. There is a singularity theorem which shows the inevitability of the formation of the spacetime singularity under some circumstances. However, there exists no such theory on the existence of an event horizon around the singularity. In 1969, Rodger Penrose gave a cosmic censorship conjecture (CCC) which does not allow horizon-less strong spacetime singularity  \cite{penrose}. However, there are series of literature on the continual gravitational collapse of
the inhomogeneous matter cloud where it is shown that spacetime singularities formed during gravitational collapse can be visible by the outside observer\cite{joshi,goswami,mosani1,mosani2,mosani3,mosani4}.
Therefore, the spacetime singularity may or may not be visible by asymptotic observer, since it depends upon the initial conditions of the collapsing matter \cite{Joshi:2011zm}. In the regime of general relativity, such a scenario can arise where the continual gravitational collapse may lead to an equilibrium static spacetime which has a central visible or naked singularity e.g., Joshi-Malafarina-Narayan (JMN), Janis-Newman-Winicour (JNW), and Bertrand (BST) spacetimes \cite{Joshi:2011zm,JNW,perlick}.

From the above discussion, it is clear now that the naked singularity can be formed as the end state of continuous gravitational collapse of an inhomogeneous matter cloud. Therefore, if naked singularity really exits in nature, it would have distinguishable physical signatures. Since every compact object in our universe has its intrinsic angular momentum, for more realistic scenario, one needs to construct rotating counterpart of the naked singularity spacetime. In this paper, we first construct the rotating version of Janis-Newman-Winicour (JNW) naked singularity spacetime to investigate its possible physical signatures in the context of shadow shape and the dynamics of timelike bound orbits.
One can obtain the rotating naked singularity spacetime by applying the NJA to the static naked singularity metric. 

However, when we apply the NJA to other spacetime metrics, such as naked singularities and other black hole solutions, the final result which is obtained in the Eddington-Finkelstein coordinates (EFC) may not be transformed into the BLC due to the complexification procedure as shown in \cite{Mustapha2011}. 
It is shown by Mustapha Azreg-Aïnou that the rotating spacetime generated by the NJA and written in the EFC could be transformed into the BLC by dropping the complexification step \cite{Mustapha1, Mustapha2, Mustapha3}. 
Recently, the rotating form of the polytropic black hole spacetime is constructed using the NJA without complexification \cite{polytropic}. 
The use of NJA is described in other theories of gravity as well, i.e., f(R) gravity, Einstein-Maxwell-Dilaton gravity, Born-Infeld monopole gravity \cite{Canonico}. 
To illustrate this method, in this paper, we apply the NJA without complexification to the static Janis-Newman-Winicour (JNW) naked singularity spacetime. Using the procedure given in \cite{Mustapha1, Mustapha2, Mustapha3}, we obtain the rotating JNW naked singularity spacetime  and it does satisfy all the energy conditions. 

As it is stated above, the black hole and naked singularity should have some different physical signatures, since the geometrical properties of them are different from each other. Therefore, observational investigations on different physical properties of ultra compact objects may reveal the nature of the same. 
  The Event Horizon Telescope (EHT) collaboration recently released the first-ever image of the shadow of the black hole located at the center of Messier 87 (M87) galaxy \cite{Kazunori_Akiyama}. 
  Similar shadows can also be cast by other compact objects. There are several literature where the shadows cast by compact objects such as black holes, naked singularities, gravastar and wormholes are extensively studied \cite{Gralla:2019xty,Abdikamalov:2019ztb,Vagnozzi:2019apd,Gyulchev:2019tvk,Dey:2013yga,Dey+15,Dey:2020haf,atamurotov_2015, abdujabbarov_2015b, ohgami_2015, stuchlik_2019, Kaur, Sakai, Shaikh:2021,Li:2021,Hu:2020usx}. 
  In \cite{Shaikh:2018lcc}, authors discuss the shadow cast by Joshi-Malafarina-Narayan (JMN) naked singularity spacetime and they compare the results with a shadow cast by the Schwarzschild black hole. In the first type of JMN  (JMN1) naked singularity spacetime, one can define a range of characteristic parameter ($0<M_0<2/3$) for which there exists no photon sphere and hence no shadow and therefore, an interesting full-moon image of the central singularity is formed.
  However, for $M_0>2/3$, JMN1 naked singularity can cast a similar shadow what an equally massive  Schwarzschild black hole can cast. 
It is generally believed that the shadows arise due to the presence of a photon sphere. Recently, in \cite{Joshi2020}, the authors have introduced a new spherically symmetric naked singularity solution of the Einstein field equation which does not have a photon sphere, but the singularity still casts a shadow. After that, in \cite{Dey:2020bgo}, for null-like and timelike naked singularities, the general conditions for the shadow formation in the absence of a photon sphere are investigated. 

The shape of the shadows and their properties would be different for the rotating spacetimes. The involvement of a spin parameter can make a spacetime more realistic, and the shadow cast by that spacetime can give us a more realistic shadow shape of a compact object. In \cite{Shaikh:2019fpu}, a general formula has been derived to obtain the shape of a shadow cast by a compact object whose gravitational field is described by a rotating spacetime geometry. Also, this general formula is applied to some known black hole solutions, and the corresponding results for shadows are reproduced. In our recent work \cite{parth6}, we analyse the shape of the shadows in the deformed Kerr spacetime. We show that due to the deformation, shadow with oblate and prolate shape can exist for the negative and positive sign of the deformation parameter respectively. In last few years, the shadows cast by rotating black holes are extensively studied in many literature \cite{abdujabbarov_2015a,abdujabbarov_2017,younsi_2016,papnoi_2014,Afrin,Ghosh:2020ece,Chen:2020aix,Haroon:2018ryd,Ovgun:2018tua,Lee:2021sws,Konoplya:2021,Shaikh:2021yux}. 
The shadow in the rotating naked singularity may have a different distinguishable signature. Therefore, in this paper, we study the shadow shape in rotating JNW spacetime and we show that the shape of the shadows in rotating JNW spacetime can be distinguished from the shadows cast by the Kerr black hole.

On the other hand, the GRAVITY, SINFONI, and UCLA galactic center groups are continuously observing astrometric and spectroscopic data of the S-stars which are orbiting around our Milky-way galactic center \cite{Do:2019txf,GRAVITY:2018ofz,Hees:2017aal}. These observational data could help to reveal the nature of the galactic center (Sgr-A*). 
The relativistic orbital precession in the Schwarzschild spacetime is discussed in \cite{GRAVITY:2020gka}. There are large number of literature where the nature of the orbit precession is extensively studied in various spacetime geometry \cite{Martinez, Eva,Eva1,Eva2,tsirulev,Joshi:2019rdo,Bambhaniya:2019pbr,Dey:2019fpv,Bam2020,Lin:2021noq,Deng:2020yfm,Deng:2020hxw,Gao:2020wjz}. In \cite{Joshi:2019rdo,Bambhaniya:2019pbr}, the periastron precession of particles orbit in static JMN1, JMN2, JNW, and Bertrand (BST) naked singularity spacetimes are studied, and the results obtained are compared with that of the Schwarzschild spacetime. It is concluded that the nature of orbital precession in BST is similar to that in Schwarzschild spacetime which is in the same direction as the motion of a particle. On the contrary, an orbit can precess in the opposite direction of the motion of a particle in JMN1, JMN2, and JNW spacetimes.

Similarly, as we discussed above, the rotating spacetimes are more physically realistic to investigate the orbital precession. 
In our previous work \cite{Bam2020, parth6}, we have shown that the orbits in deformed Kerr spacetime can precess in the opposite direction of the motion of a particle when the deformed Kerr black hole becomes a naked singularity, which is not possible in Kerr spacetime. 
We have also shown that the shadow and negative precession of timelike bound orbits can simultaneously exist in deformed Kerr naked singularity spacetime. There are some literature where the timelike orbits in rotating black holes are extensively studied \cite{aa4,Glampedakis:2002ya,Fujita:2009bp,Pugliese:2013zma,rana}. Here, we analyse the precession of the timelike bound orbits in rotating JNW spacetime and compare them with the Kerr black hole case. We find that the rotating JNW spacetime can admit negative precession for the particular range of scalar field charge $q$ and spin parameter $a$. These novel features of the shadow shape and the precession of the timelike bound orbits in rotating JNW spacetime can be observationally significant to distinguish from the Kerr and other rotating compact objects.  

This paper is constructed in the following way. In section (\ref{Sec_Rotating_JNW_NJA}), we derive the rotating JNW naked singularity spacetime using the NJA with and without complexification. We discuss the energy conditions in both cases. In section (\ref{Sec_Shadow}), we study the shadow properties and find the shape of the shadows in rotating JNW spacetime. In addition, we show the comparison of the shape of the shadow cast by the Kerr black hole and rotating JNW naked singularity. In section (\ref{Sec_Orbit}), we obtain the general differential orbit equation for the timelike bound orbit and compare the orbital precession in the Kerr and rotating JNW spacetime. Finally, we discuss and conclude our result in (\ref{sec_discussion}). We consider gravitational constant $G$ and speed of light $c$ equal to unity throughout this paper.

\section{Construction of the rotating JNW spacetime metric}
\label{Sec_Rotating_JNW_NJA}

In this section, we determine the rotating JNW naked singularity spacetime by using the Newman-Janis algorithm (NJA). The final result of the NJA is obtained in the null coordinates, which must be transformed into the Boyer-Lindquist coordinates (BLC). The BLCs are more convenient to work with the rotating spacetime metric. We analyse that it is not possible to perform the BLC transformation when considering JNW spacetime metric (\ref{JNW_metric}) as a seed metric. To resolve this problem, we drop the complexification step in NJA \cite{Mustapha1, Mustapha2, Mustapha3}.

\subsection{JNW metric and the Newman Janis Algorithm}
The JNW spacetime metric is given by
\begin{eqnarray}
    ds^2 = -\left(1-\frac{2M}{r\nu} \right)^{\nu} dt^2 + \left(1-\frac{2M}{r\nu} \right)^{-\nu} dr^2 \nonumber \\ +\, r^2 \left(1-\frac{2M}{r\nu} \right)^{1-\nu} d\Omega^2
    \label{JNW_metric}
\end{eqnarray}
where, $d\Omega^2 = d\theta^2 + \sin^2\theta d\phi^2$. It is straight forward to show that the following expression is the final form of the NJA with considering JNW metric (\ref{JNW_metric}) as a seed metric.
\begin{eqnarray}
    ds^2 = \, && - \left(1-\frac{2Mr}{\nu \rho^2} \right)^{\nu} du^2 - 2 du dr \nonumber \\ && -\, 2a\sin^2\theta \left[1 - \left(1 - \frac{2Mr}{\nu \rho^2} \right)^{\nu} \right] du d\Phi \nonumber \\ && +\, \rho^2 \left(1 - \frac{2Mr}{\nu \rho^2} \right)^{1-\nu} d\Omega^2 + 2a\sin^2\theta dr d\Phi \nonumber \\ && +\, a^2\sin^4\theta \left[2 - \left(1-\frac{2Mr}{\nu \rho^2} \right)^{\nu} \right] d\Phi^2
    \label{JNW_metric_EFC}
\end{eqnarray}
The above form of the spacetime metric seems rather complicated. It can be made more simple and symmetrical by transforming it into the Boyer-Lindquist coordinates (BLC). The BLC $(t, r, \theta, \phi)$ represents the rotating black hole spacetime metrics because all the off-diagonal terms of the metric, except $dt d\Phi$, vanish; and its axial symmetry becomes apparent. The transformation into the BLC requires
\begin{eqnarray}
    && du = dt - \xi(r) dr
    \label{du}\\
    && d\Phi = d\phi - \chi(r) dr
    \label{dphi}
\end{eqnarray}
Here, we note that the transformation functions, i.e., $\xi(r)$ and $\chi(r)$, strictly depends on coordinate $r$ only; otherwise eq. (\ref{du}) and (\ref{dphi}) will not be integrable. To write eq. (\ref{JNW_metric_EFC}) into the BLC like form, the transformation functions must take the following form.
\begin{eqnarray}
    && \xi(r,\theta) = \frac{1}{\Delta} \left[\rho^2 \left(1 - \frac{2Mr}{\nu \rho^2} \right)^{1-\nu} + a^2\sin^2\theta \right]\\
    && \chi(r) = \frac{a}{\Delta}\\
    && \Delta = r^2 - \frac{2Mr}{\nu} + a^2
\end{eqnarray}
It can be seen that the transformation function $\xi$ also depends on coordinate $\theta$, and therefore the eq. (\ref{du}) is not integrable. In these coordinates, eq. (\ref{JNW_metric_EFC}) is written as
\begin{widetext}
\begin{eqnarray}
    ds^2 = -\left(1-\frac{2Mr}{\nu \rho^2} \right)^{\nu} dt^2 + \rho^2 \left(1 - \frac{2Mr}{\nu \rho^2}\right)^{1-\nu} \left(\frac{dr^2}{\Delta} + d\Omega^2 \right) - 2a\sin^2\theta \left[1 - \left(1 - \frac{2Mr}{\nu \rho^2} \right)^{\nu} \right] dt d\phi \nonumber \\+ a^2\sin^4\theta \left[2 - \left(1 - \frac{2Mr}{\nu \rho^2} \right)^{\nu} \right] d\phi^2
    \label{Rotating_JNW_NJA}
\end{eqnarray}
\end{widetext}
It can be seen that the above metric reduces to the JNW spacetime metric (\ref{JNW_metric}) when $a=0$. Although the transformation into the BLC was improper in obtaining eq. (\ref{Rotating_JNW_NJA}), we can check the energy conditions because the NJA is just an algorithm that successfully works for Kerr and Kerr-Newman metric. Thus, we cannot say whether the final result eq. (\ref{Rotating_JNW_NJA}) is a valid solution of the Einstein field equation. It can only be answered once we know whether the energy conditions for the metric (\ref{Rotating_JNW_NJA}) are satisfied or not. The energy conditions for the above spacetime metric are discussed in the Appendix, and we conclude that the rotating JNW spacetime metric (\ref{Rotating_JNW_NJA}) obtained using NJA with complexification violates the energy conditions.\\
Now, we drop the $\theta$ dependency of the transformation function $\xi$ by considering a slow rotation approximation in which the second and higher-order terms of rotation parameter ($a$) are ignored. Thus, eq. (\ref{Rotating_JNW_NJA}) reduces to the following form.
\begin{eqnarray}
    ds^2 = -\left(1-\frac{2M}{\nu r} \right)^{\nu} dt^2 + \left(1-\frac{2M}{\nu r} \right)^{-\nu} dr^2 \nonumber \\ -\, 2a\sin^2\theta \left[1 - \left(1-\frac{2M}{\nu r}\right)^{\nu} \right] dt d\phi \nonumber \\ +\, r^2 \left(1-\frac{2M}{\nu r} \right)^{1-\nu} d\Omega^2
    \label{Slow_rotation_JNW}
\end{eqnarray}
And, the transformation functions $\chi$ and $\xi$ become
\begin{eqnarray}
     && \xi(r) = \left(1-\frac{2M}{\nu r} \right)^{-\nu} \nonumber \\
    && \chi(r) = \frac{a}{r^2} \left(1-\frac{2M}{\nu r} \right)^{-1} \nonumber
\end{eqnarray}
Eq. (\ref{Slow_rotation_JNW}) is the slow rotation form of the JNW spacetime metric. We note that the rotating JNW spacetime metric (\ref{Rotating_JNW_NJA}) obtained using the NJA with complexification is not properly transformed into the BLC because of the complexification of the coordinates. This issue can be solved by skipping the complexification step \cite{Mustapha3,polytropic}, which is discussed in the next section.

\subsection{Newman-Janis Algorithm without Complexification}

\noindent Consider a general form of the static, and spherically symmetric spacetime metric.
\begin{equation}
ds^2=-G(r)dt^2+\frac{dr^2}{F(r)}+H(r)d\Omega^2
\label{general_metric}
\end{equation}
To apply the algorithm, transform it into the null coordinates $\{u,r,\theta,\phi\}$ using the coordinate transformation
\begin{equation}
du=dt-\frac{dr}{\sqrt{F(r)G(r)}}
\end{equation}
In null coordinates, the spacetime metric (\ref{general_metric}) takes the form
\begin{equation}
ds^2=-G(r)du^2-2\sqrt{\frac{G(r)}{F(r)}}dudr+H(r)d\Omega^2
\label{general_metric_null}
\end{equation}
The null tetrad of the above spacetime metric are
\begin{eqnarray}
&& l^\mu=\delta^\mu_1\,, \label{l_mu} \\
&& n^\mu=\sqrt{\frac{F(r)}{G(r)}}\delta^\mu_0-\frac{1}{2}F(r)\delta^\mu_1\,, \label{n_mu} \\
&& m^\mu=\frac{1}{\sqrt{2H(r)}} \left(\delta^\mu_2+\frac{i}{\sin\theta}\delta^\mu_3 \right) \label{m_mu}
\end{eqnarray}
The next step of the NJA is the complexification, which generalize the function (i.e. $G(r), F(r), H(r)$) as the real function of radial coordinate $r$ and it's complex conjugate $\bar{r}$. The trick is to skip this step and proceed directly to the next step which is the complex coordinate transformation.
\begin{eqnarray}
&& r^\prime=r+ia\cos\theta \label{r_prime} \\
&& u^\prime=u-ia\cos\theta \label{u_prime}
\end{eqnarray}
Under the coordinate transformation, let us consider that the components of the metric tensor (\ref{general_metric}) are transformed as $$\{G(r),F(r),H(r)\}\to\{A(r,\theta,a),B(r,\theta,a),\psi(r,\theta,a)\},$$ where the following conditions are satisfied.
\begin{eqnarray}
&& \lim\limits_{a\to 0}A(r,\theta,a)=G(r)\,, \\
&& \lim\limits_{a\to 0}B(r,\theta,a)=F(r)\,, \\
&& \lim\limits_{a\to 0}\psi(r,\theta,a)=H(r).
\end{eqnarray}
Now, perform the complex coordinate transformation (\ref{r_prime}), (\ref{u_prime}) on the null tetrad (\ref{l_mu}), (\ref{n_mu}), (\ref{m_mu}).
\begin{eqnarray}
&& l^\mu=\delta^\mu_1\,, \\
&& n^\mu=\sqrt{\frac{B}{A}}\delta^\mu_0-\frac{1}{2}B\delta^{\mu}_1\,, \\
&& m^\mu=\frac{1}{\sqrt{2\psi}} \left(\delta^\mu_2+ia\sin\theta (\delta^\mu_0-\delta^\mu_1) +\frac{i}{\sin\theta}\delta^\mu_3 \right)
\end{eqnarray}
The forth null vector $\bar{m}^\mu$ is just a complex conjugate of $m^\mu$. These are the components of the null tetrad of the final metric tensor in null coordinates $(u, r, \theta, \Phi)$.
\begin{eqnarray}
    ds^2 = A du^2+2\sqrt{\frac{A}{B}}dudr+2a\sin^2\theta\left(\sqrt{\frac{A}{B}}-A \right)dud\Phi \nonumber \\ -\, 2a\sin^2\theta\sqrt{\frac{A}{B}}drd\Phi-\psi d\theta^2 \nonumber \\ -\, \sin^2\theta\left[\psi+a^2\sin^2\theta\left(2\sqrt{\frac{A}{B}}-A \right)\right]d\Phi^2 \,\,\,\,\,\,\,\,\,\,\,
    \label{general_rotating_null}
\end{eqnarray}
The spacetime metric is determined in the null coordinates $\{u,r,\theta,\Phi\}$, which can be transformed into the BLC $\{t,r,\theta,\phi\}$ as
\begin{eqnarray}
&& du=dt-\lambda(r)dr
\label{du new}\\
&& d\Phi=d\phi-\chi(r)dr
\label{dphi new}
\end{eqnarray}
The transformation functions $\chi$ and $\lambda$ should strictly depend on the radial coordinate ($r$) only, for eq. (\ref{du new}), (\ref{dphi new}) to be integrable; which is not possible in NJA with complexification. As we know, in Boyer-Lindquist form, the spacetime metric contains only one off-diagonal term which is $dtd\phi$. Therefore, using the above coordinate transformation (\ref{du new}, \ref{dphi new}), and dropping all off-diagonal terms except $dtd\phi$, we have the following two expressions.
\begin{widetext}
    \begin{eqnarray}
        && A\lambda(r)-\sqrt{\frac{A}{B}}+a\sin^2\theta\left(\sqrt{\frac{A}{B}}-A \right)\chi(r)=0
        \label{dtdr}\\
        && a\left(\sqrt{\frac{A}{B}}-A \right)\lambda(r)+a\sqrt{\frac{A}{B}}+\left[\psi-a^2\sin^2\theta\left(2\sqrt{\frac{A}{B}}-A \right)\chi(r) \right]=0
        \label{drdphi}
    \end{eqnarray}
\end{widetext}
Solving these expressions for $\lambda$ and $\chi$, we obtain
\begin{eqnarray}
&& \chi(r,\theta)=\frac{a}{\psi B+a^2\sin^2\theta}\,, \\
&& \lambda(r,\theta)=\frac{\psi\sqrt{\frac{B}{A}}+a^2\sin^2\theta}{\psi B+a^2\sin^2\theta}
\end{eqnarray}
Here, again, the BLC transformation functions $\chi$ and $\lambda$ depends on $\theta$; but this time we have three unknown functions $A(r,\theta,a), B(r,\theta,a)$ and $\psi(r,\theta,a)$ which play an important role in dropping the $\theta$ dependency of $\chi$ and $\lambda$. We choose the following form of the transformation functions, because it successfully works in Kerr and Kerr-Newman case.
\begin{eqnarray}
&& \chi(r)=\frac{a}{FH+a^2}
\label{chi}\\
&& \lambda(r)=\frac{K+a^2}{FH+a^2}
\label{lambda}
\end{eqnarray}
Where, $K=H\sqrt{\frac{F}{G}}$. \\
Now, substitute the above expression of $\chi$ and $\lambda$ into the equations ($\ref{dtdr},\ref{drdphi}$), and solve them for the unknown functions $A(r,\theta,a)$ and $B(r,\theta,a)$. We obtain
\begin{eqnarray}
&& A(r,\theta,a)=\frac{(FH+a^2\cos^2\theta)\psi}{(K+a^2\cos^2\theta)^2}\\
&& B(r,\theta,a)=\frac{FH+a^2\cos^2\theta}{\psi}
\end{eqnarray}
We can verify that the function $A(r,\theta,a)$ and $B(r,\theta,a)$ respectively reduce to the function $G(r)$ and $F(r)$ when $a\to 0$.
\begin{eqnarray}
&& \lim\limits_{a\to 0}A(r,\theta,a)=G(r)\\
&& \lim\limits_{a\to 0}B(r,\theta,a)=F(r)
\end{eqnarray}
Therefore, under the BLC coordinate transformation, the metric (\ref{general_rotating_null}) becomes
\begin{widetext}
    \begin{eqnarray}
         ds^2=-\frac{(FH+a^2\cos^2\theta)\psi}{(K+a^2\cos^2\theta)^2}dt^2+\frac{\psi}{FH+a^2}dr^2-2a\sin^2\theta\left(\frac{K-FH}{(K+a^2\cos^2\theta)^2} \right)\psi dtd\phi+\psi d\theta^2 \nonumber \\ +\, \psi sin^2\theta\left[1+a^2\sin^2\theta\left(\frac{2K-FH+a^2\cos^2\theta}{(K+a^2\cos^2\theta)^2} \right) \right]d\phi^2
        \label{general_rotating_BLC}
    \end{eqnarray}
\end{widetext}
where, $\psi(r,\theta,a)$ is still unknown. It must satisfy the following constraint which corresponds to the vanishing component of the Einstein tensor, $G_{r\theta}=0$.
\begin{equation}
    3a^2\sin{2\theta}\psi^2 K_{,r} + (K+a^2\cos^2\theta)^2 (3 \psi_{,\theta}\psi_{,r} - 2 \psi \psi_{,r\theta}) = 0
\end{equation}
This non-linear partial differential equation can be solved for $\psi$. One can verify that a solution of the above partial differential equation is
\begin{equation}
    \psi(r,\theta,a) = K(r) + a^2\cos^2\theta
    \label{psi}
\end{equation}
For the Kerr and Kerr-Newman metric, $K(r)=r^2$, and the above expression reduces to $\psi=r^2+a^2\cos^2\theta$. After substituting $\psi$ (\ref{psi}) into the general form of the rotating spacetime metric (\ref{general_rotating_BLC}), if the static spacetime metric (\ref{general_metric}) can be obtained in the limit $a\to 0$, the rotating solution is known as normal fluid; otherwise it is known as conformal fluid. Normal fluid solution requires
\begin{equation}
    \lim\limits_{a\to 0}\psi(r,\theta,a)=H(r)
\end{equation}
which implies $F(r)=G(r)$.\\
The metric (\ref{general_rotating_BLC}) can be written in Kerr-like form as
\begin{eqnarray}
    ds^2 = -\left(1-\frac{2f}{\rho^2} \right)dt^2 + \frac{\rho^2}{\Delta} dr^2 + \rho^2 d\theta^2 + \frac{\Sigma\sin^2\theta}{\rho^2} d\phi^2 \nonumber \\ -\,  \frac{4af\sin^2\theta}{\rho^2} dt d\phi \,\,\,\,\,\,\,\,\,\,
    \label{general_rotating_metric}
\end{eqnarray}
Where,
\begin{eqnarray}
    && \rho^2 = K(r) + a^2\cos^2\theta\,, \\
    && f = \frac{K(r)-F(r)H(r)}{2}\,, \\
    && \Delta = F(r) H(r) + a^2\,, \\
    && \Sigma = \big(K(r)+a^2 \big)^2 - a^2\Delta\sin^2\theta.
\end{eqnarray}
When $a\to 0$, the rotating solution (\ref{general_rotating_metric}) reduces to the static, and spherically symmetric spacetime metric if $F(r) = G(r)$.

\subsection{Rotating JNW spacetime metric}
We construct a new rotating spacetime metric which reduces to the static JNW naked singularity spacetime metric (\ref{JNW_metric}) in the limit $a\to 0$. Substitute the components of the JNW spacetime metric into eq. (\ref{general_rotating_metric}).
\begin{eqnarray}
    ds^2 = -\left(1-\frac{2f}{\rho^2} \right)dt^2 + \frac{\rho^2}{\Delta} dr^2 + \rho^2 d\theta^2 + \frac{\Sigma\sin^2\theta}{\rho^2} d\phi^2 \nonumber \\ - \frac{4af\sin^2\theta}{\rho^2} dt d\phi \,\,\,\,\,\,\,\,\,\,\,\,
    \label{JNW_rotating}
\end{eqnarray}
Where,
\begin{eqnarray}
    && 2f = r^2 \left(1 - \frac{2M}{r\nu} \right) \left[-1 + \left(1 - \frac{2M}{r\nu} \right)^{-\nu} \right]\,, \\
    && \rho^2 = r^2 \left(1 - \frac{2M}{r\nu} \right)^{1-\nu} + a^2\cos^2\theta\,, \\
    && \Delta = r^2 - \frac{2Mr}{\nu} + a^2\,, \\
    && \Sigma = (\rho^2 + a^2\sin^2\theta)^2 - a^2\Delta\sin^2\theta.
\end{eqnarray}
One can verify that it reduces to the Kerr metric when $\nu = 1$. Also it reduces to eq. (\ref{Slow_rotation_JNW}) under the slow rotation approximation. To check whether the spacetime metric is physically valid, we must proceed to check the energy conditions. Consider the inverse form of the rotating spacetime metric (\ref{JNW_rotating}).
\begin{eqnarray}
    (\partial_s)^2 = - \frac{(K+a^2)^2-a^2\Delta\sin^2\theta}{\rho^2\Delta} (\partial_t)^2 + \frac{\Delta}{\rho^2} (\partial_r)^2 \nonumber \\ +\, \frac{1}{\rho^2} (\partial_\theta)^2 + \frac{\Delta-a^2\sin^2\theta}{\rho^2\Delta\sin^2\theta} (\partial_\phi)^2 - \frac{4 a M K}{r\rho^2\Delta} \partial_t\partial_\phi \,\,\,
\end{eqnarray}
It can also be written as
\begin{eqnarray}
   (\partial_s)^2 = -\left(\frac{K+a^2}{\rho\sqrt{\Delta}}\partial_t + \frac{a}{\rho\sqrt{\Delta}}\partial_{\phi} \right)^2 + \frac{\Delta}{\rho^2} (\partial_r)^2 \nonumber \\
   +\, \frac{1}{\rho^2} (\partial_{\theta})^2 + \left(\frac{a\sin\theta}{\rho}\partial_t + \frac{1}{\rho\sin\theta}\partial_{\phi} \right)^2 \,\,\,
\label{Inverse_general_rotating_metric}
\end{eqnarray}
Now, we choose a set of orthonormal basis $\{u^{\mu}, e_{r}^{\mu}, e_{\theta}^{\mu}, e_{\phi}^{\mu} \}$, where $u^\mu$ can be considered as a four velocity of the fluid.
\begin{eqnarray}
    && u^\mu = \left(\frac{K+a^2}{\rho\sqrt{\Delta}}, 0, 0, \frac{a}{\rho\sqrt{\Delta}} \right)\,, \\
    && e_r^\mu = \left(0, \frac{\sqrt{\Delta}}{\rho}, 0, 0 \right)\,, \\
    && e_\theta^\mu = \left(0, 0, \frac{1}{\rho}, 0 \right)\,, \\
    && e_\phi^\mu = \left(\frac{a\sin\theta}{\rho}, 0, 0, \frac{1}{\rho\sin\theta} \right).
\end{eqnarray}
One can verify that $u^\mu u_\mu = -1$, $e_{i}^{\mu} (e_i)_\mu = 1$, $u^\mu (e_i)_\mu = 0$; where, $(i \to r,\theta,\phi)$. In terms of these orthonormal basis, eq. (\ref{Inverse_general_rotating_metric}) is written as
\begin{equation}
    (\partial_s)^2 = (-u^\mu u^\nu + e_r^\mu e_r^\nu + e_\theta^\mu e_\theta^\nu + e_\phi^\mu e_\phi^\nu) \partial_\mu \partial_\nu
\end{equation}
where the metric tensor is
\begin{equation}
    g^{\mu\nu} = -u^\mu u^\nu + e_r^\mu e_r^\nu + e_\theta^\mu e_\theta^\nu + e_\phi^\mu e_\phi^\nu
\end{equation}
Similarly, the energy-momentum tensor can be decomposed as
\begin{equation}
    T^{\mu\nu} = \rho_e u^\mu u^\nu + P_r e_r^\mu e_r^\nu + P_\theta e_\theta^\mu e_\theta^\nu + P_\phi e_\phi^\mu e_\phi^\nu
\end{equation}
where, $\rho_e$ is the energy density, and $P_i$ $(i\to r,\theta,\phi)$ are the principal pressure components. The Einstein field equation in the form $G_{\mu\nu}=T_{\mu\nu}$ requires
\begin{eqnarray}
    && \rho_e = u^\mu u^\nu G_{\mu\nu} \label{rho general}\,, \\
    && P_r = e_r^\mu e_r^\nu G_{\mu\nu} = g^{rr} G_{rr} \label{Pr general}\,, \\
    && P_\theta = e_\theta^\mu e_\theta^\nu G_{\mu\nu} = g^{\theta\theta} G_{\theta\theta} \label{Ptheta general}\,, \\
    && P_\phi = e_\phi^\mu e_\phi^\nu G_{\mu\nu} \label{Pphi general}.
\end{eqnarray}
The components of the energy-momentum tensor of the rotating JNW spacetime are as follows. We consider $\theta = \pi/2$ for the mathematical simplicity.
\begin{eqnarray}
    && \rho_e = \frac{M^2 (1 - \nu^2) (\Delta+a^2)}{r^6 \nu^2} \left(1 - \frac{2M}{r\nu} \right)^{-3+\nu} \label{rho} \\
    && P_r = - P_\theta = \frac{M^2 (1-\nu^2)}{r^4 \nu^2} \left(1 - \frac{2M}{r\nu} \right)^{-2+\nu} \label{Pr_Ptheta}
\end{eqnarray}
\begin{eqnarray}
    P_\phi = P_\theta + \frac{2a^2}{r^4} \left(1-\frac{2M}{r\nu} \right)^{-3+\nu} \bigg[\left(1-\frac{2M}{r\nu} \right)^{1+\nu} \nonumber \\ - \left(1-\frac{M(1+\nu)}{r\nu}\right)^2 \bigg]\,
    \label{Pphi}
\end{eqnarray}
It can be verified that the rotating JNW spacetime metric (\ref{JNW_rotating}) satisfies the weak energy condition ($\rho_e \geq 0$, $\rho_e+P_i \geq 0$); null energy condition ($\rho_e+P_i \geq 0$); strong energy condition ($\rho_e+\sum P_i \geq 0$, $\rho_e+P_i \geq 0$); and the dominant energy condition ($\rho_e > |P_i|$). We also note that when $\nu=1$, all the components eq. (\ref{rho}), (\ref{Pr_Ptheta}), (\ref{Pphi}) vanish, which is obvious as we have already seen that the rotating JNW metric reduces to the Kerr metric when $\nu=1$. Also, $P_r \neq P_\theta \neq P_\phi$, therefore, it is an an-isotropic fluid solution.

\begin{figure*}
\centering
\subfigure[Shadow of rotating JNW for $q =0.4$ and $a = 0.3$.]
{\includegraphics[width=55mm]{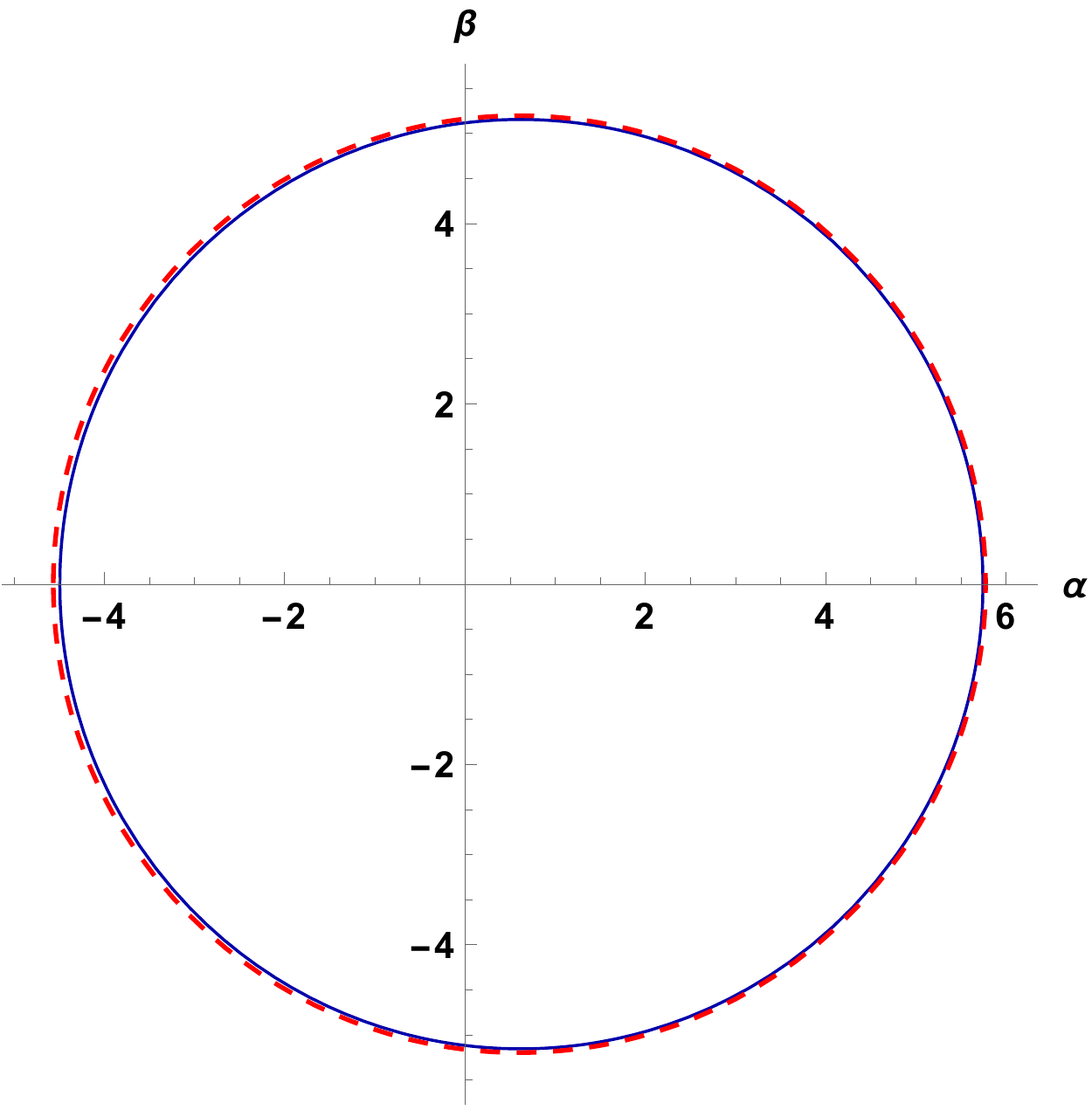}\label{re1n}}
\hspace{0.2cm}
\subfigure[Shadow of rotating JNW for $q =0.4$ and $a = 0.4$.]
{\includegraphics[width=55mm]{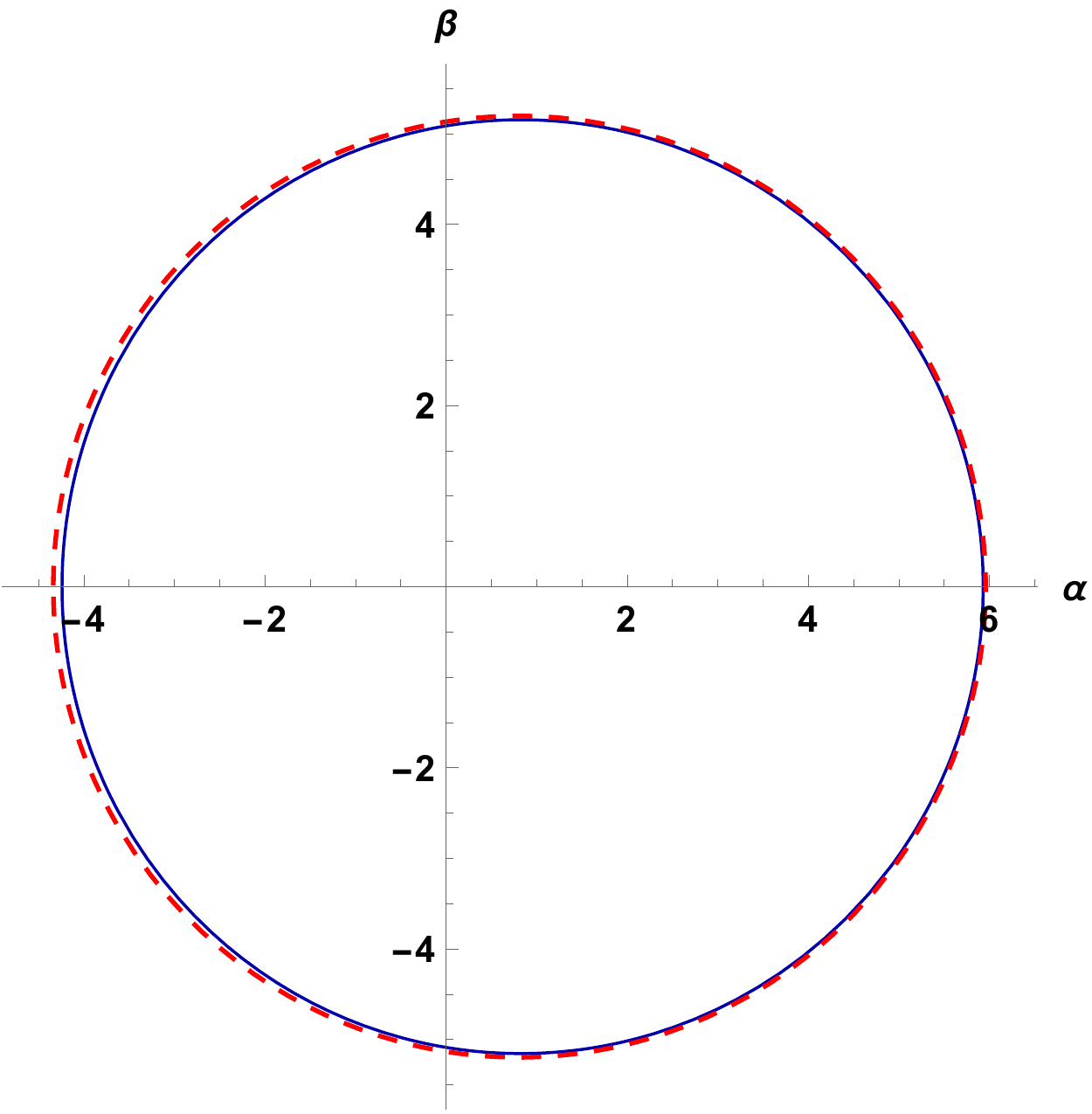}\label{re2n}}
\subfigure[Shadow of rotating JNW for $q =0.4$ and $a = 0.5$.]
{\includegraphics[width=55mm]{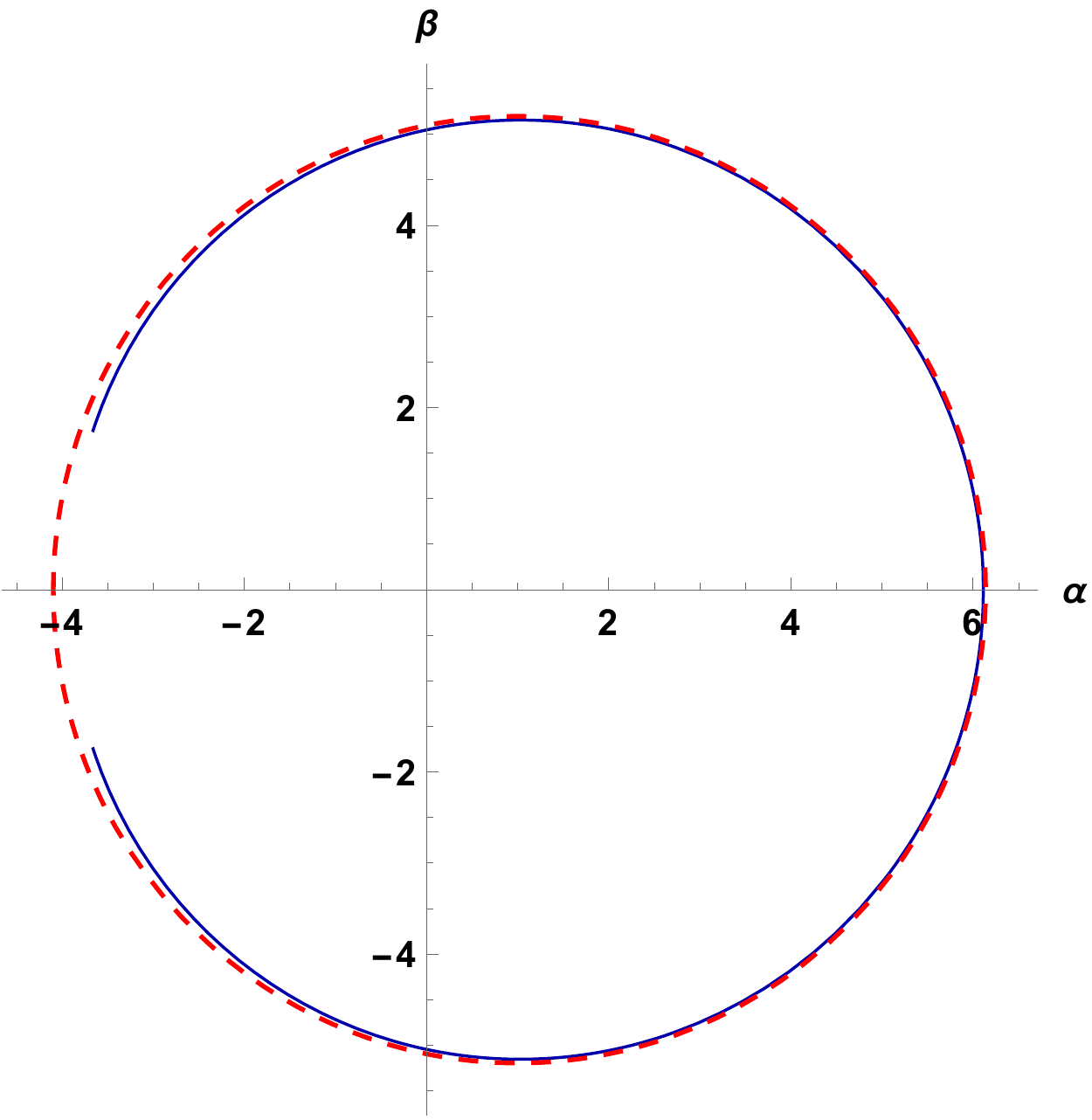}\label{re3n}}\\
\subfigure[Shadow of rotating JNW for $q =0.6$ and $a = 0.3$.]
{\includegraphics[width=55mm]{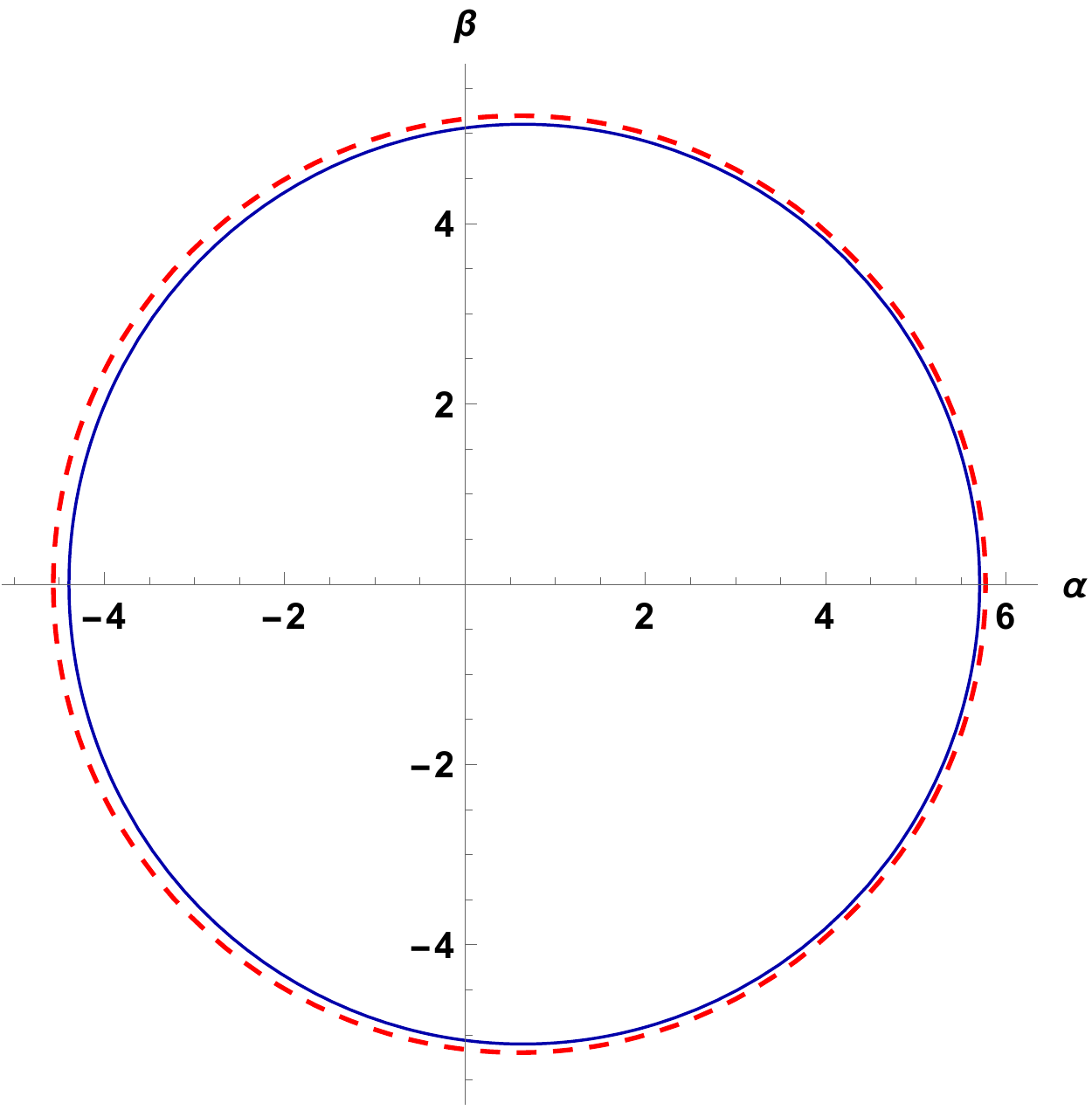}\label{re1p}}
\hspace{0.2cm}
\subfigure[Shadow of rotating JNW for $q =0.6$ and $a = 0.4$.]
{\includegraphics[width=55mm]{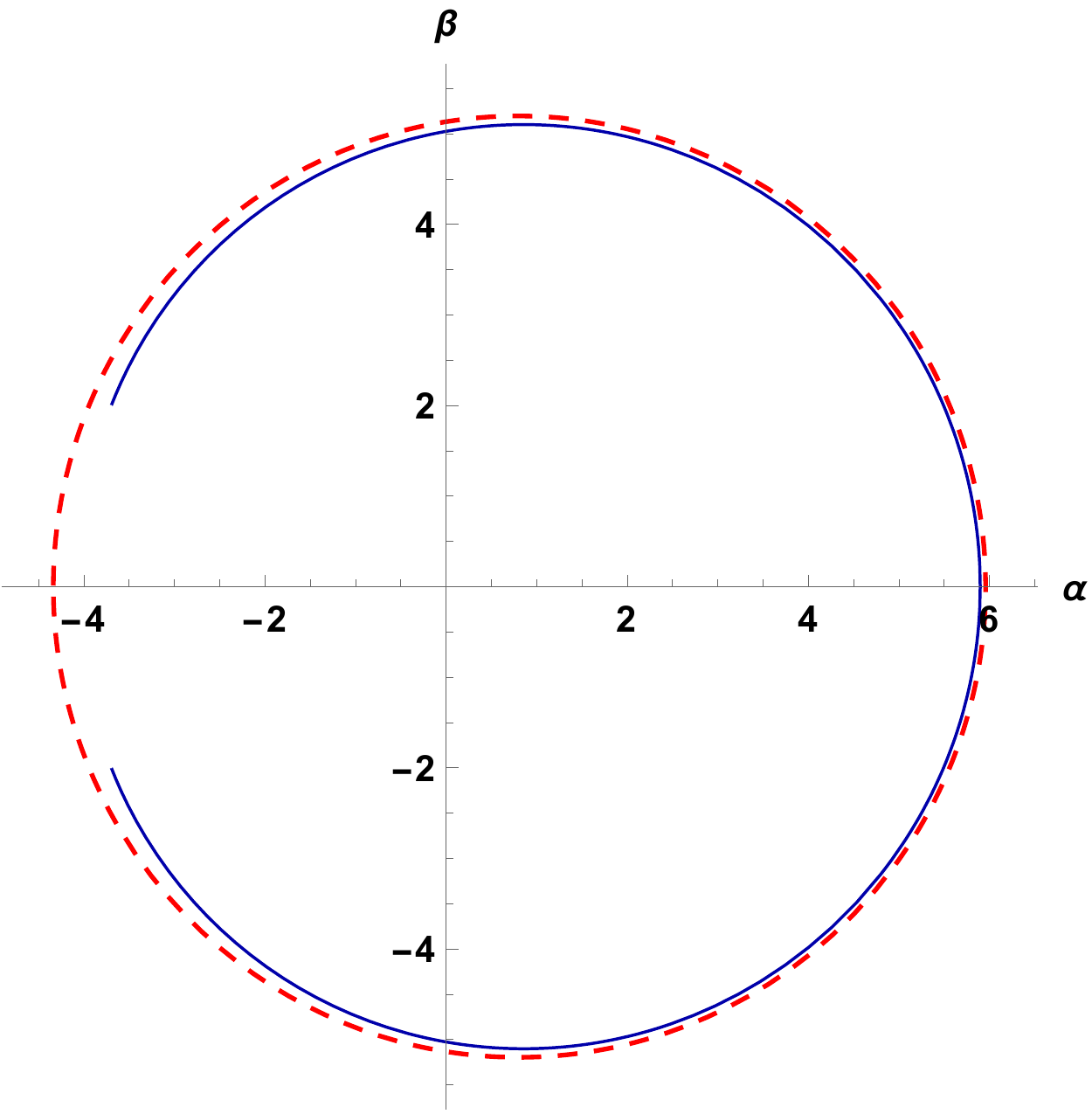}\label{re2p}}
\subfigure[Shadow of rotating JNW for $q =0.6$ and $a = 0.5$.]
{\includegraphics[width=55mm]{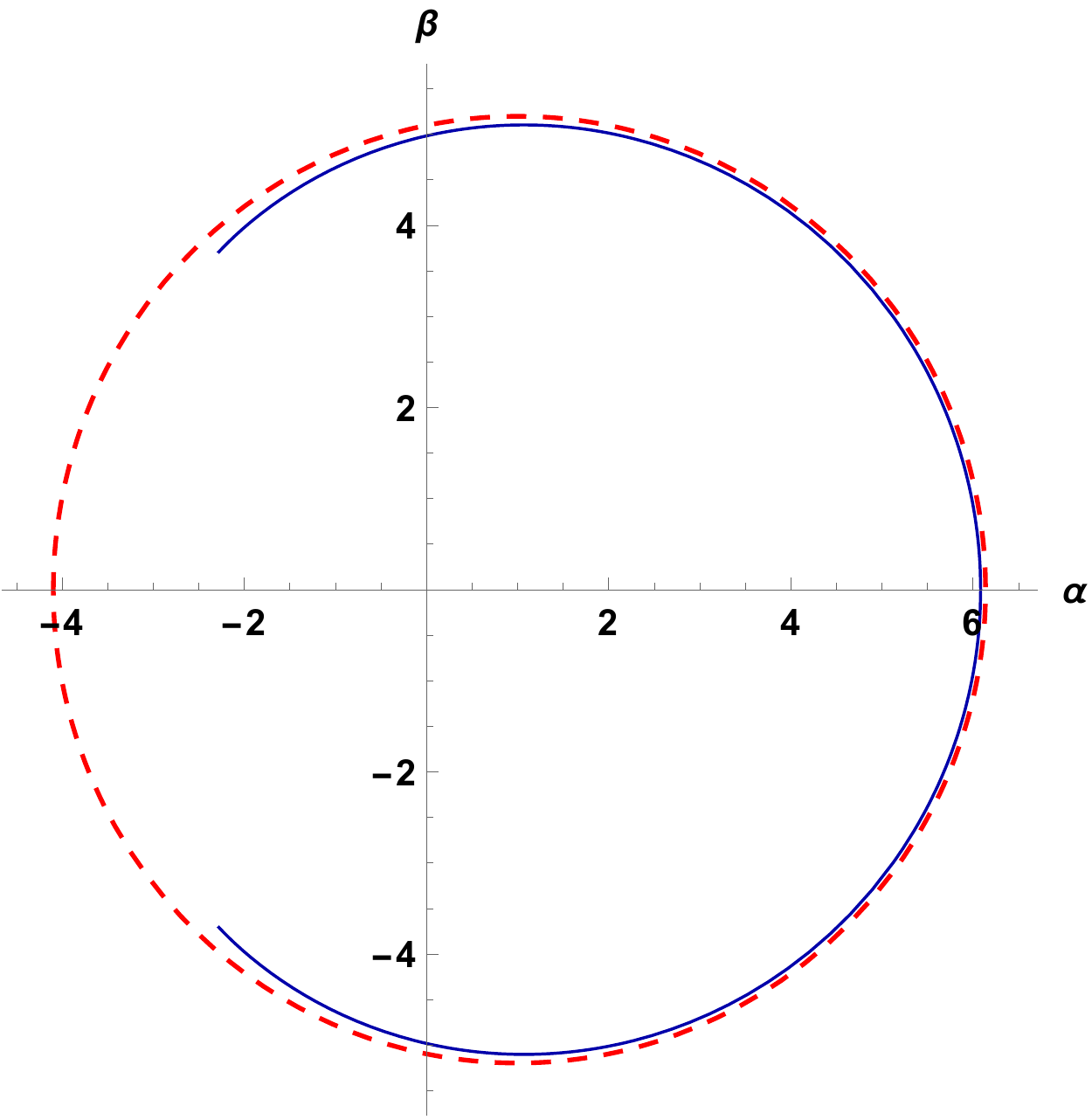}\label{re3p}}\\
\subfigure[Shadow of rotating JNW for $q =0.8$ and $a = 0.3$.]
{\includegraphics[width=55mm]{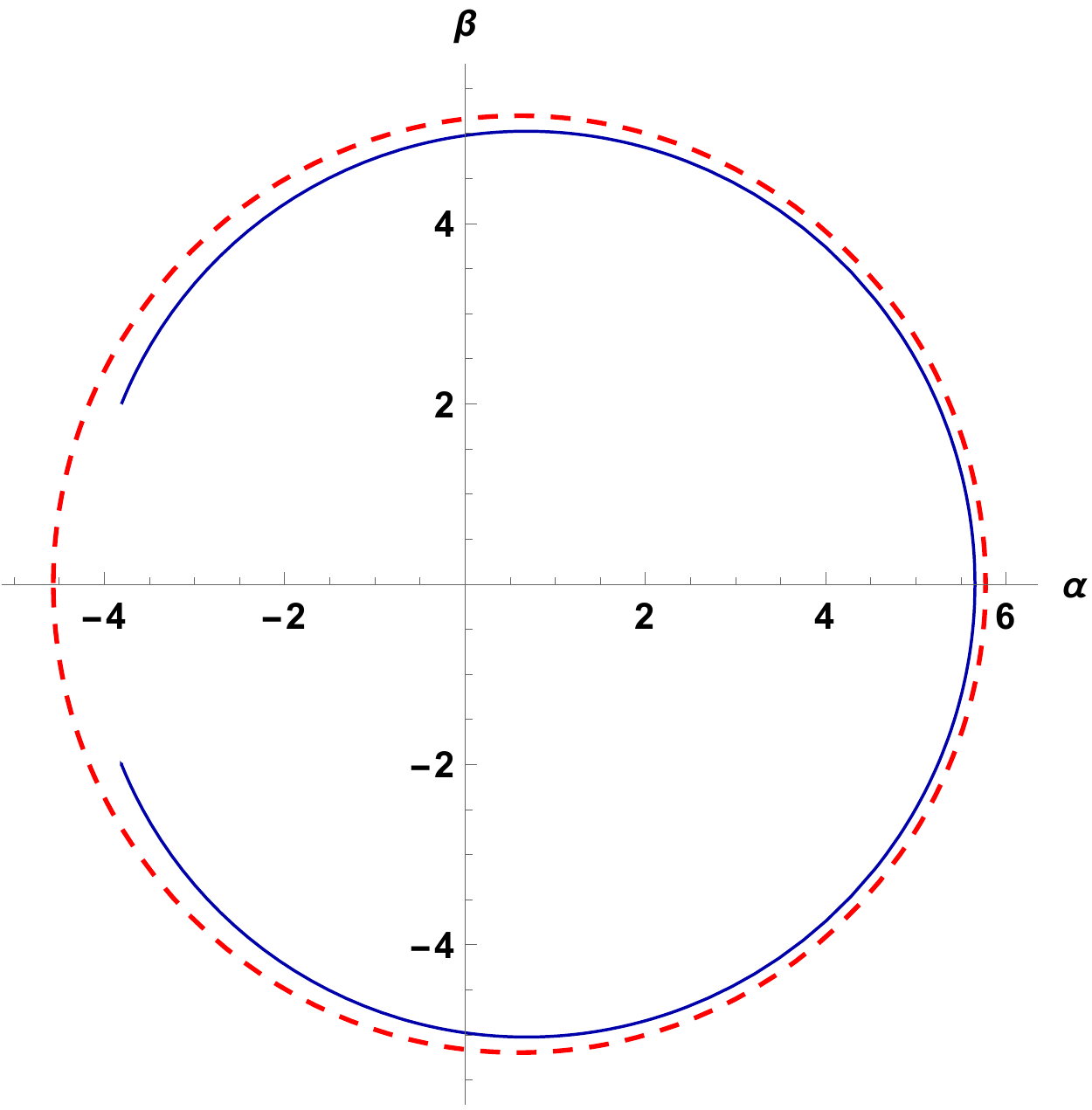}\label{re11p}}
\hspace{0.2cm}
\subfigure[Shadow of rotating JNW for $q =0.8$ and $a = 0.4$.]
{\includegraphics[width=55mm]{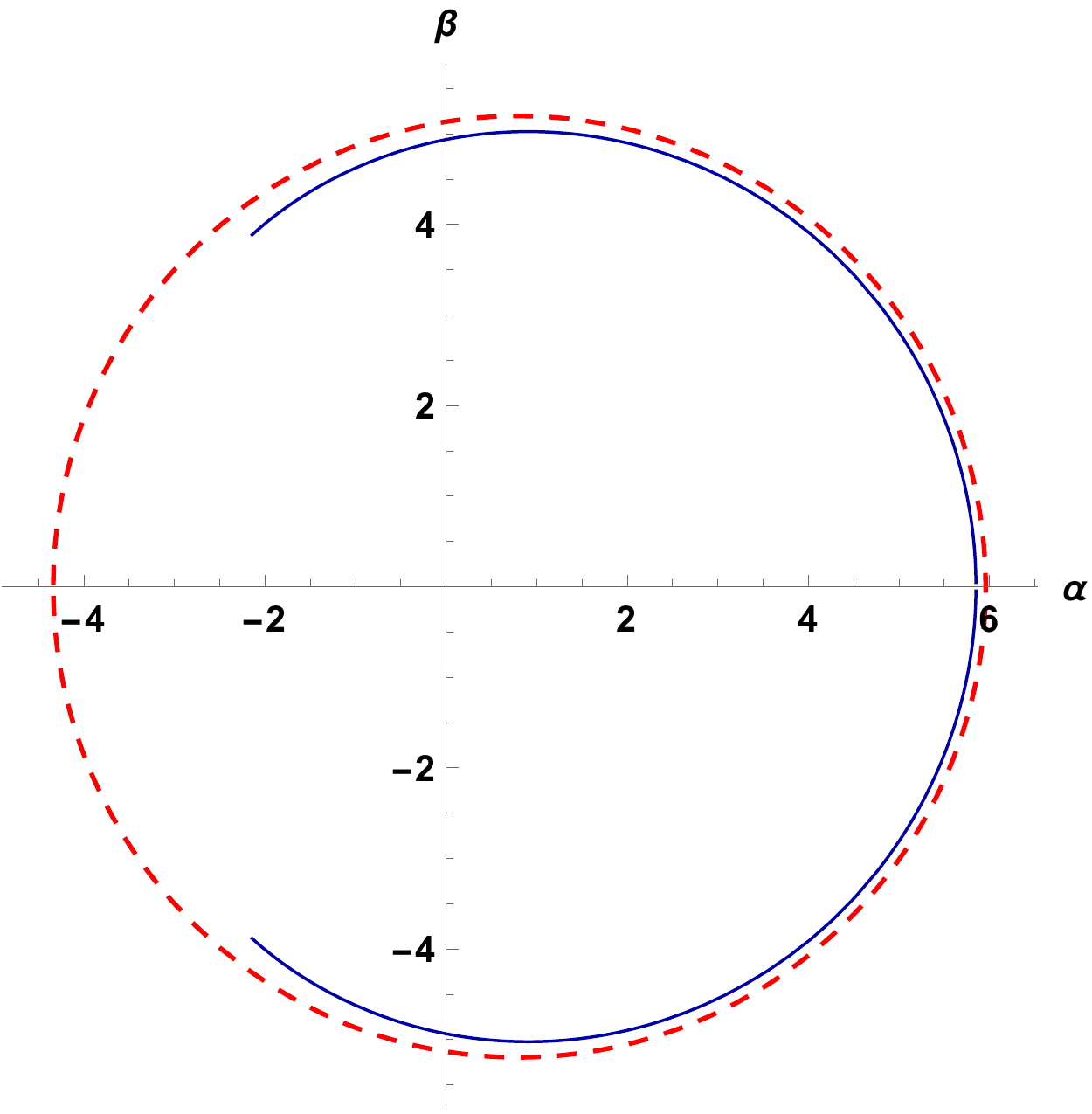}\label{re22p}}
\subfigure[Shadow of rotating JNW for $q =0.8$ and $a = 0.5$.]
{\includegraphics[width=55mm]{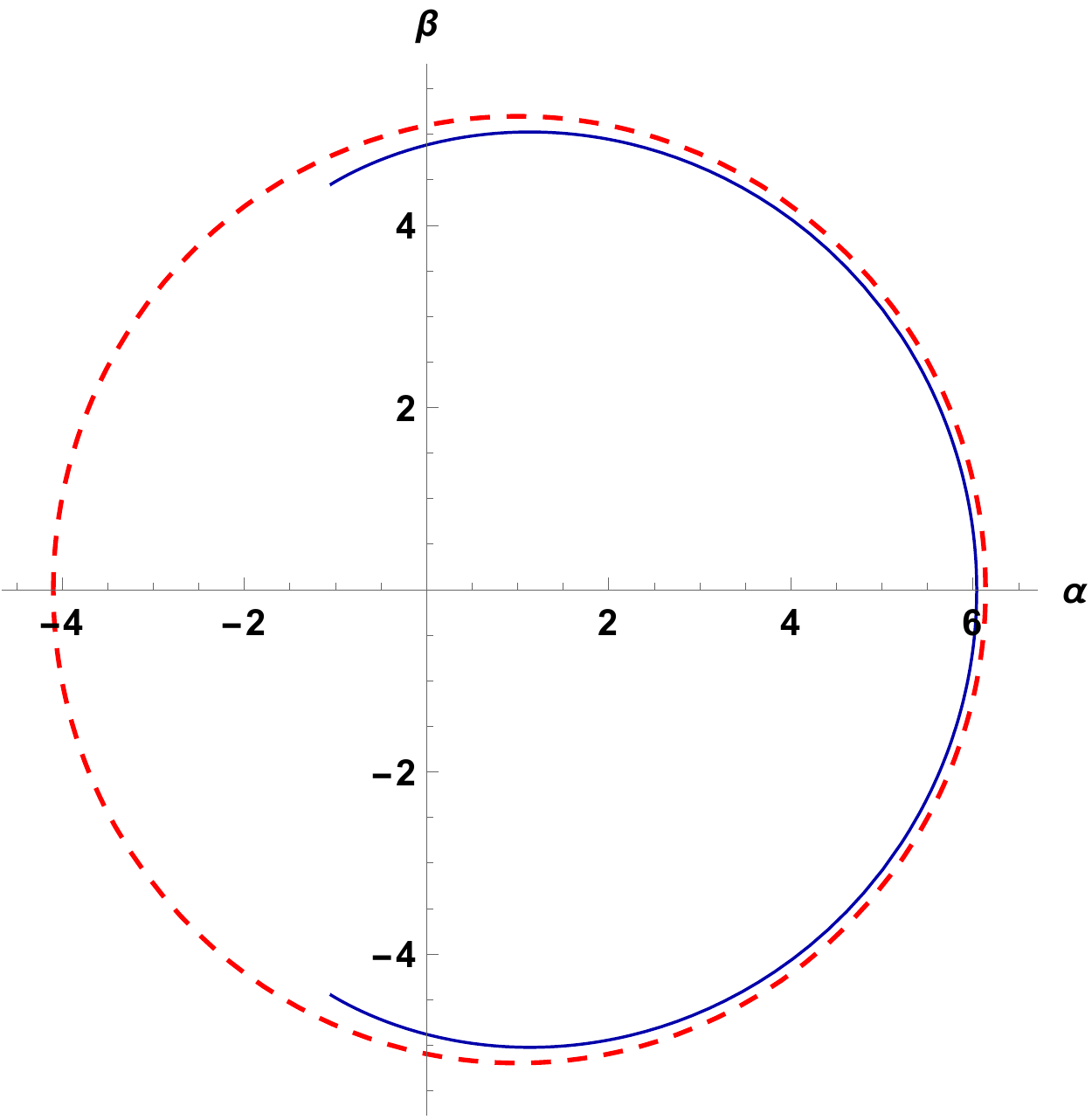}\label{re33p}}\\
 \caption{Figure shows the shape of the shadow cast by the rotating JNW naked singularity for various values of scalar field charge $q$ and spin parameter $a$. Here, we consider $M=1$ and $\theta=\pi/2$ for the shape of shadows in celestial coordinates. The details of this figure are discussed in the text.}
\label{figshadow}
\end{figure*}

\section{Shadows in rotating JNW spacetime}
\label{Sec_Shadow}

In this section, we review the general formalism of the null geodesic using the Hamilton-Jacobi separation method and obtained the general expressions for finding the shape and nature of a shadow. In general relativity, the Hamilton-Jacobi equation is given by,
\begin{equation}
   \frac{\partial S}{\partial \lambda}+\mathcal{H} = 0, \;
   \mathcal{H}=\frac{1}{2} g^{\mu \nu}p_{\mu}p_{\nu}
   \label{Hamilton},
\end{equation}
where $\lambda$ is an affine parameter, S is the Jacobi action, $\mathcal{H}$ is the Hamiltonian, and $p_
{\mu}$ is the momentum
defined by,
\begin{equation}
   p_
{\mu}=\frac{\partial S}{\partial x^{\mu}}=g_{\mu \nu}\frac{d x^{\nu}}{d\lambda}.
\end{equation}
In the above equation (\ref{Hamilton}), the Hamiltonian does not depend explicitly on $t$ and $\phi$ coordinates. Therefore we have two constants of motion, the conserved energy $E=-p_t$ and the conserved angular momentum
$L=p_\phi$ (about the axis of symmetry).
Since we have the separable solution of the above differential equation (\ref{Hamilton}), the Jacobi action can be written in terms of
already known constants of the motion as,
\begin{equation}
   S = \frac{1}{2}\mu^2 \lambda - E t + L\phi + S_{r}(r) + S_{\theta}(\theta) \label{action},
\end{equation}
 where $\mu$ is the rest mass of the test particle. Therefore, for a photon or null geodesic, we consider $\mu=0$. The metric tensor components of general rotating spacetime (\ref{general_rotating_BLC}) are given as, 

 \begin{equation}
   g^{tt}=-\frac{H-a^2 G \sin^2\theta+2 a^2 \sqrt{\frac{G}{F}}\sin^2\theta}{P}, 
 \end{equation}
  \begin{equation}
   g^{rr}=\frac{FH+a^2\sin^2\theta}{H},
 \end{equation}
 \begin{equation}
   g^{\theta\theta}=\frac{1}{H},
 \end{equation}
 \begin{equation}
   g^{\phi\phi}=-\frac{G\csc^4\theta}{2a^2G^2-4a^2G\sqrt{\frac{G}{F}}+a^2\left(\frac{G}{F}\right)-GH\csc^2\theta},
 \end{equation}
 \begin{equation}
   g^{t\phi}=g^{\phi t}=\frac{a\left( \sqrt{\frac{G}{F}}-G\right)}{P}, 
 \end{equation}
where,
\begin{widetext}
\begin{equation}
    P=-GH+2a^2G^2\sin^2\theta-4a^2G\sqrt{\frac{G}{F}}\sin^2\theta+a^2 \left(\frac{G}{F}\right)\sin^2\theta
\end{equation}
\end{widetext}
Substituting eq. (\ref{action}) into eq. (\ref{Hamilton}), and using the components of the metric tensor, we obtain the following expression:
\begin{widetext}
\begin{equation}
    -\left(FH+a^2\sin^2\theta\right)\left(\frac{dS_r}{dr}\right)^2+\frac{\left(\left(\sqrt{\frac{F}{G}}H+a^2\sin^2\theta\right)E-aL\right)^2}{FH+a^2\sin^2\theta}-\left(L-aE\right)^2=\left(\frac{dS_{\theta}}{d\theta}\right)^2+L^2\cot^2\theta-a^2E^2\cos^2\theta,
    \label{p1}
\end{equation}
\end{widetext}
Note that, left and right hand side of above eq. (\ref{p1}) are only functions of $r$ and $\theta$ respectively, therefore one can write each side equal to a separation constant since Jacobi action principle is separable solution,
\begin{widetext}
\begin{equation}
  -\left(FH+a^2\sin^2\theta\right)\left(\frac{dS_r}{dr}\right)^2+\frac{\left(\left(\sqrt{\frac{F}{G}}H+a^2\sin^2\theta\right)E-aL\right)^2}{FH+a^2\sin^2\theta}-\left(L-aE\right)^2=\mathcal{K},  
\end{equation}

\begin{equation}
  \left(\frac{dS_{\theta}}{d\theta}\right)^2+L^2\cot^2\theta-a^2E^2\cos^2\theta=\mathcal{K},  
\end{equation}
\end{widetext}
where the separation constant $\mathcal{K}$ is known as the Carter constant. Using above differential equations one can write down the following separated null geodesic equations for radial ($r$) and tangential ($\theta$) part as,
\begin{equation}
    H^2\left( \frac{dr}{d\lambda}\right)^2 -R(r) = 0\,\, ,
    \label{rmotion}
\end{equation}
\begin{equation}
    H^2\left( \frac{d\theta}{d\lambda}\right)^2 -\Theta(\theta) = 0\,\, ,
    \label{thetamotion}
\end{equation}
where, $R(r)$ and $\Theta(\theta)$ can be considered as radial and tangential part of effective potentials respectively, which can be defines as,
\begin{equation}
   R(r) = \left[X(r) - a \xi \right]^2 - \Delta(r) \left[\eta + (\xi - a)^2 \right]\,\, ,
   \label{rpoten}
\end{equation}
\begin{equation}
\Theta(\theta)=\eta+(\xi-a)^2-\left(\frac{\xi}{\sin\theta}-a\sin\theta\right)^2\,\,
\label{thetapoten}
\end{equation}
where, $\xi = \frac{L}{E}$ and $\eta = \frac{\mathcal{K}}{E^2}$. The functions $X(r)$ and $\Delta(r)$ can be define as,
\begin{equation}
    X(r)=\sqrt{\frac{F}{G}}H+a^2\sin^2\theta,
\end{equation}
\begin{equation}
    \Delta(r)=FH+a^2\sin^2\theta.
\end{equation}
Note that, $R(r)$ and $\Theta(\theta)$ must be non-negative for photon motion. In a general rotating spacetime, unstable circular orbits exist when the following conditions hold,
\begin{equation}
     R(r_{ph}) = 0, \;   \frac{d R(r_{ph})}{dr}=0, \;    \frac{d^2 R(r_{ph})}{dr^2} \leq 0\,\,. \label{eq3}
\end{equation}
where, $r=r_{ph}$ is the radius of the unstable photon orbit.
Using the above conditions (\ref{eq3}), we obtain the critical impact parameters corresponding to the maxima of the $R(r)$ as,
\begin{equation}
     \xi = \frac{X_{ph}\Delta'_{ph}-2\Delta_{ph}X'_{ph}}{a\Delta'_{ph}},
     \label{xi}
\end{equation}
\begin{equation}
    \eta = \frac{4a^2X'_{ph}\Delta_{ph}-\left[(X_{ph} - a^2)\Delta'_{ph}-2X'_{ph}\Delta_{ph}\right]^2}{a^2\Delta'^2_{ph}},
    \label{eta}
\end{equation}
where prime denoted for derivative concerning $r$ and subscript ``ph" denoted for the quantities which are evaluated at $r=r_{ph}$. The above expressions are the general forms of the critical impact parameters $\xi$ and $\eta$ for the unstable photon orbits, which would be important to find out the shape of the shadows. However, one can obtain the apparent shape of the shadow by using the celestial coordinates $\alpha$ and $\beta$ which lie in the line of sight of the observer direction. The expressions of the celestial coordinates $\alpha$ and $\beta$ can be written in the following way,
\begin{eqnarray}
    \alpha &=& \lim_{r_0\to\infty} \left(-r_0^2 \sin{\theta_{0}} \frac{d \phi}{dr}\right)\,\, ,\\
    \beta &=& \lim_{r_0\to\infty} \left(r_0^2 \frac{d \theta}{dr}\right)\,\, ,
\end{eqnarray}
where $r_0$ and $\theta_0$ are the coordinates of the asymptotic observer. Note that, if we consider the general rotating metric to be an asymptotically flat in the limit of $r\to \infty$, then we can consider $G\to 1$, $F\to 1$, $H\to r^2$, $X\to r^2$ and $\Delta\to r^2$. Therefore, the above expressions of the celestial coordinates $\alpha$ and $\beta$ becomes,
\begin{eqnarray}
    \alpha &=& -\frac{\xi}{\sin{\theta_{0}}}\,\, ,\\
    \beta &=& \pm \sqrt{\eta + a^2 \cos^2{\theta_{0}} - \xi^2\cot^2{\theta_{0}}}\,\, ,
\end{eqnarray}
from the above expressions (\ref{xi}) and (\ref{eta}), one can obtain the celestial coordinates $\alpha$ and $\beta$ for the rotating JNW spacetime by using its metric tensor components. Note that, we used JNW metric tensor components without adopting any complexification method since the general rotating metric we obtained by skipping the complexification method. The metric tensor components of the JNW spacetime are given as,
\begin{equation}
    G = F = \left(1-\frac{2M}{r\nu}\right)^{\nu}\,\, ;\
    H = r^2\left(1-\frac{2M}{r\nu}\right)^{1-\nu}
\end{equation}
using above equations we can find that the expressions for $X(r)$ and $\Delta(r)$,
\begin{equation}
    X(r)=r^2\left(1-\frac{2M}{r\nu}\right)^{1-\nu}+a^2\sin^2\theta,
\end{equation}
\begin{equation}
    \Delta(r)=r^2\left(1-\frac{2M}{r\nu}\right)+a^2\sin^2\theta.
\end{equation}
therefore, using the above expressions, one can calculate the critical impact parameters and celestial coordinates for the rotating JNW spacetime.  We have constructed the shape of the shadows using the expressions of celestial coordinates for rotating JNW spacetime, where we consider the inclination angle of the observer $\theta_{0}=\pi/2$.

\begin{figure*}
\centering
\subfigure[Timelike bound orbit in rotating JNW metric for $q =0.4$ and $a = 0.3$.]
{\includegraphics[width=75mm]{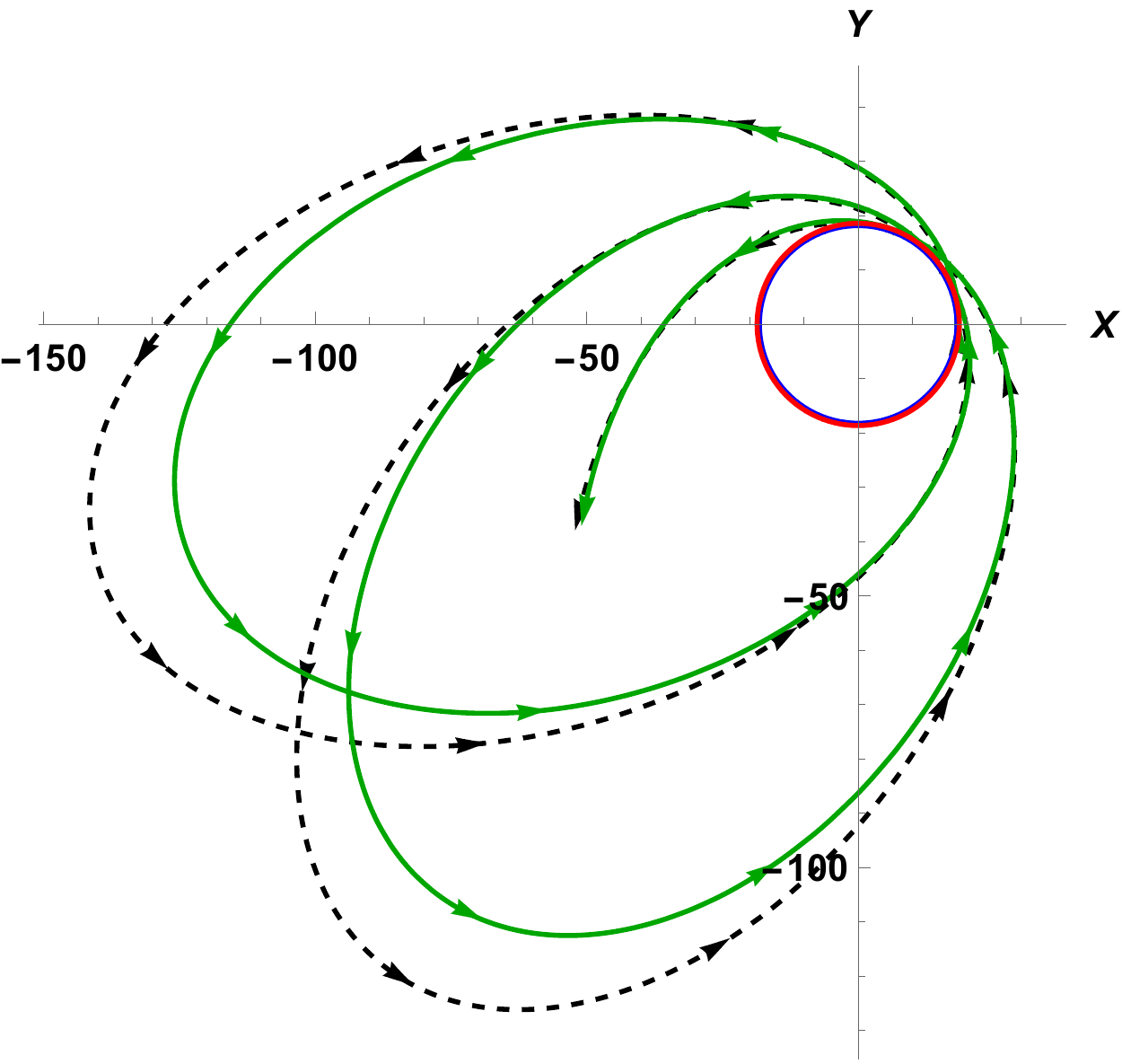}\label{orbit11}}
\hspace{0.2cm}
\subfigure[Timelike bound orbit in rotating JNW metric for $q =0.4$ and $a = 0.5$.]
{\includegraphics[width=75mm]{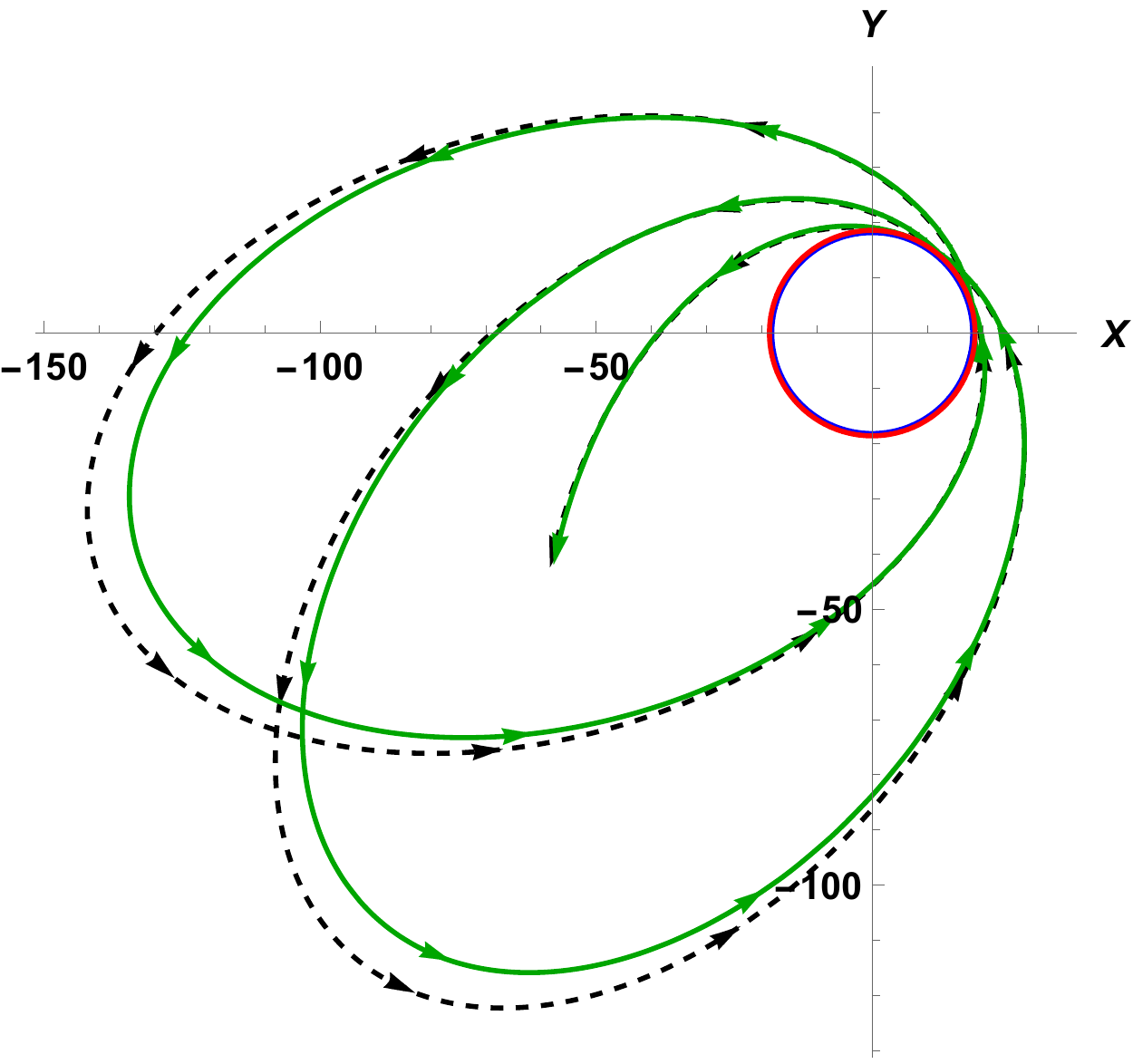}\label{orbit12}}
\subfigure[Timelike bound orbit in rotating JNW metric for $q =12$ and $a = 0.3$.]
{\includegraphics[width=75mm]{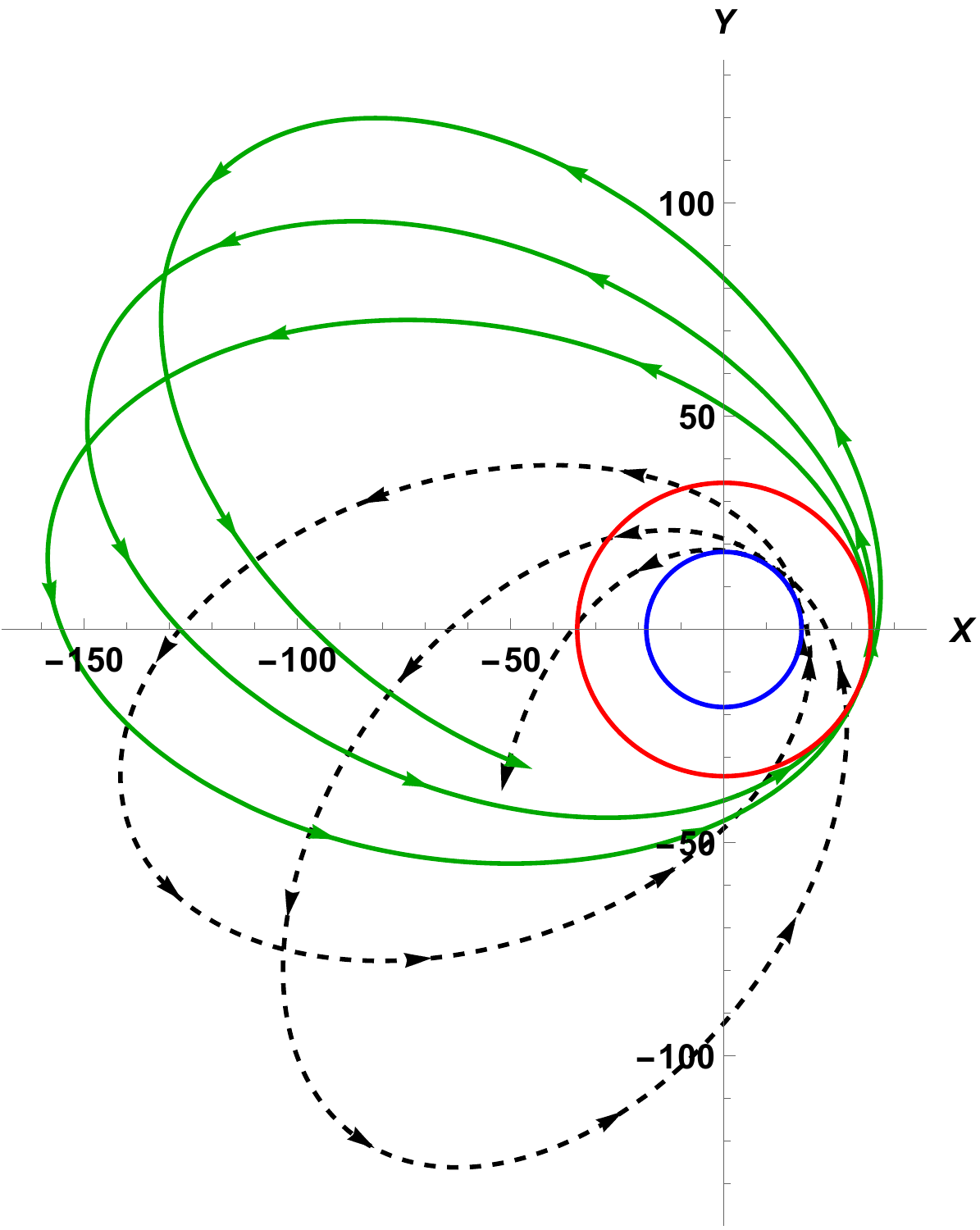}\label{orbit21}}
\subfigure[Timelike bound orbit in rotating JNW metric for $q =12$ and $a = 0.5$.]
{\includegraphics[width=75mm]{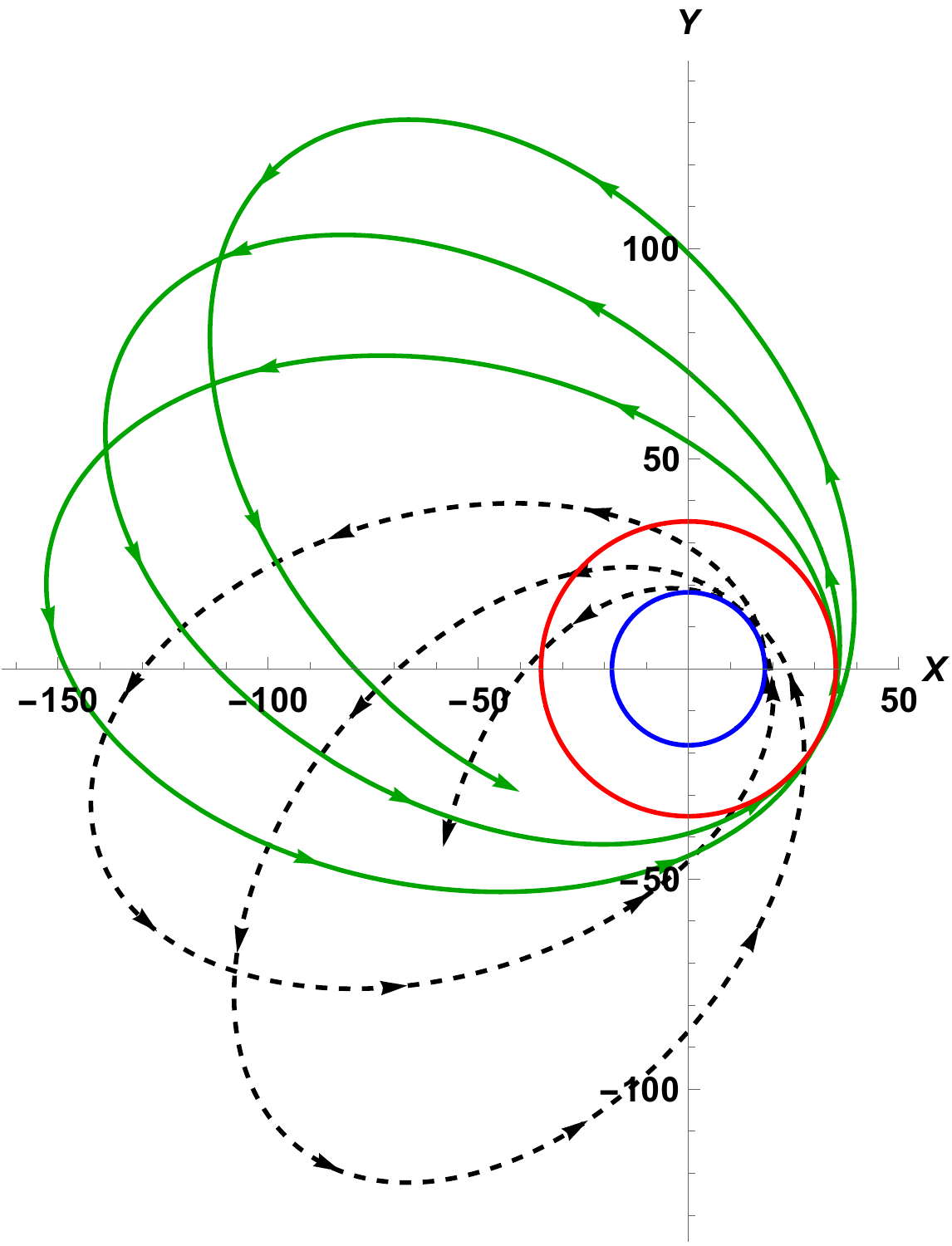}\label{orbit22}}

 \caption{The figure shows the timelike bound orbits in the rotating JNW naked singularity spacetime for scalar field charge $q=0.4, 12$ and spin parameter $a=0.3, 0.5$ respectively. Here, we consider mass (ADM mass for JNW) $M=1$, conserved angular momentum $L=6$ and total energy $E=-0.006$. Note that we consider the timelike bound orbits in the equatorial plane (i.e. $\theta=\pi/2$). The black dotted lines represent the bound orbits in the Kerr spacetime, and the solid green lines represent the bound orbits in the JNW spacetime. The corresponding minimum approach of a test particle towards the center is shown by the blue (in the Kerr metric) and the red (in the JNW metric) circles. The details of this figure are discussed in the text.}
\label{figorbit}
\end{figure*}

In fig.(\ref{figshadow}), we have shown that the shape of the shadows for different values of spin parameter $a$ and scalar field charge $q$, where the blue line indicate the shadows in rotating JNW spacetime and the red dotted line indicate the shadows in the Kerr spacetime. It is well known that the shadow shapes in the Kerr black hole spacetime always be in contour shape, and as we increase the spin parameter, it will become more and more prolate. One can see in fig.(\ref{figshadow}), the horizontal radius of the red dotted line will decreases from the left side as we increase the value of a spin parameter. On the other hand, we found that the shadows in rotating JNW spacetime changes from contour shape to arc shape as we increase the values of spin parameter and scalar field charge or one of them.

\section{Timelike geodesics in rotating JNW spacetime}
\label{Sec_Orbit}
 In this section, we derive the effective potential and the orbit equation for a test particle in rotating JNW spacetime. Solving the orbit equation numerically, we determine the nature of the orbit precession. We proceed by considering a general form of the spacetime metric, i.e.,
 \begin{equation}
     ds^2 = -g_{tt} dt^2 + g_{rr} dr^2 - 2g_{t\phi} dtd\phi + g_{\theta\theta} d\theta^2 + g_{\phi\phi} d\phi^2
     \label{general rotating metric}
 \end{equation}
 which is written in the Boyer-Lindquist Coordinates (BLC). For the above spacetime metric, the effective potential takes the form
\begin{equation}
    V_{eff}=\frac{1}{2 g_{rr}} + \frac{e^2-1}{2} + \frac{L^2 g_{tt}+2eL g_{t\phi}-e^2 g_{\phi\phi}}{2g_{rr}(g_{t\phi}^2+g_{tt}g_{\phi\phi})}
\label{Veff general}
\end{equation}
where, we have considered a case of equatorial orbits for the mathematical simplicity. Substituting the metric components and taking $\theta=\frac{\pi}{2}$ into eq. (\ref{Veff general}), we obtain the effective potential for the orbits in equatorial plane of rotating JNW spacetime.
 \begin{eqnarray}
     V_{eff} = \frac{(L-ae)^2}{2r^2} \left(1-\frac{2M}{r\nu} \right)^{2\nu-1} - \frac{1}{2} \nonumber \\ +\, \frac{\Delta+2ae\left(L-ae \right)}{2r^2} \left(1-\frac{2M}{r\nu} \right)^{\nu-1}
     \label{Veff rotating JNW}
 \end{eqnarray}
 Now, we proceed to derive an orbit equation. For the general rotating spacetime metric (\ref{general rotating metric}), we obtain \cite{parth6}
\begin{eqnarray}
\left(\frac{dr}{d\phi}\right)^2=-\frac{(g_{t\phi})^2+g_{tt}g_{\phi\phi}}{g_{rr}(e g_{t\phi} + L g_{tt})^2} [(g_{t\phi})^2&-&e^2 g_{\phi\phi}+2eL g_{t\phi}\nonumber\\+&g_{tt}&(L^2+g_{\phi\phi})] \,\,\,
\label{general orbit}
\end{eqnarray}
 where, $\theta=\frac{\pi}{2}$. Now, using the differential relation $\frac{d^2r}{d\phi^2}=\frac{1}{2\frac{dr}{d\phi}}\frac{d}{d\phi}\left(\frac{dr}{d\phi}\right)^2.$, we can determine the orbit equation as,
\begin{equation}
    \frac{d^2u(\phi)}{d\phi^2} = -\frac{1}{r^2} \frac{d^2r}{d\phi^2} + \frac{2}{r^3} \left(\frac{dr}{d\phi}\right)^2
    \label{orbiteq}
\end{equation} 
 where, $u=\frac{1}{r}$. The above eq. (\ref{orbiteq}) is the non-linear second order differential equation, thus it is very difficult to find the complete analytic solution of this equation. Hence, we can solve this orbit equation numerically for the rotating JNW metric. 

In figure (\ref{figorbit}), we show the timelike bound orbits in rotating JNW spacetime and compare them with the orbits in Kerr spacetime. Note that, for simplicity, we define the particle motion in the equatorial plane (i.e. $\theta=\pi/2$). In fig. (\ref{orbit11}),(\ref{orbit12}), (\ref{orbit21}) and (\ref{orbit22}), we show the bound orbits for scalar field charge $q=0.4, 12$ corresponding to the spin parameter $a=0.3, 0.5$. The green and black dotted lines represent the orbits of the particles in the rotating JNW and Kerr spacetimes respectively. The red and blue circles denote the periastron points (minimum approach towards the center) for rotating JNW and Kerr spacetime respectively. The parameters we have consider the mass (ADM mass for JNW) $M=1$, conserved angular momentum $L=6$, total energy $E=-0.006$.

One can see from the figure (\ref{figorbit}), the timelike bound orbital motion in the JNW spacetime is significantly different from the orbital motion in Kerr spacetime. We find that the timelike bound orbits can have negative (or opposite) precession in rotating JNW spacetime. In other words, if the angular distance traveled by a particle between two successive periastron points is less than $2\pi$, then the next orbit would shift in the opposite direction of the particle's orbiting direction. Hence, it is known as negative (or opposite) precession. On the other hand, in \cite{Bam2020}, it is shown that the negative precession is not present in the Kerr spacetime. The Kerr spacetime exhibits positive precession since the angular distance traveled by a particle between two successive periastron points is always greater than $2\pi$.  
 
The spherically symmetric and static JNW spacetime can cast the shadow due to the existence of the photon sphere \cite{Dey:2020haf}. The photon sphere exist for the range of parameter $0.5\leq \nu < 1$ and absent for the range of $0 < \nu < 0.5$. On the other hand, the negative precession of timelike bound orbits is present \cite{Bambhaniya:2019pbr} only within the range $0 < \nu < 0.1$. Therefore, in \cite{Dey:2020haf}, it is shown that the negative precession of the timelike bound orbits and shadow formation can not occur simultaneously. Similarly in the rotating JNW spacetime, the parametric range slightly differs as we increase the spin parameter $a$.     
 
 \section{Discussions and conclusions}
 \label{sec_discussion}
 The conclusions of this paper are in the following way.
\begin{itemize}

 \item We construct the rotating JNW spacetime using the NJA without employing the complexification of coordinates. We have shown that the rotating JNW spacetime obtained using the NJA with considering the complexification step cannot be written into the BLC. If we anyhow write that metric into the BLC-like form by performing an improper coordinate transformation, we find that it does not satisfy the energy conditions. To resolve this problem, we drop the complexification step in the NJA so that the rotating solution can properly be transformed into the BLC. As a result, we obtain the rotating JNW spacetime into the BLC, which does satisfy the energy conditions.
 
\item As we know, the rotating Kerr black hole can cast prolate contour shape shadow (see dotted red line in fig.(\ref{figshadow})). While we increase the spin parameter up to $M\ge a$, it will become more and more prolate contour, where $M$ is the mass of the black hole. But, as we reach the $M<a$ limit, the Kerr black hole becomes the Kerr naked singularity and it would not cast any shadow. To get more insight, in this paper, we obtain the shape of the shadows cast by rotating JNW naked singularity and compare them with the shape of shadows cast by the Kerr black hole. We find that the shape of the shadows in rotating JNW spacetime becomes prolate arc shape from the prolate contour shape as we increase the scalar field charge $q$ and/or spin parameter $a$ (see blue line in fig. (\ref{figshadow})). 

\item  In \cite{Bambhaniya:2019pbr,Bam2020}, we have shown that the Schwarzschild and Kerr black holes admit only positive orbital precession. In other words, timelike bound orbits only precess in the direction of particle's motion. Here, we have examined the precession of timelike bound orbits in rotating JNW spacetime. We find that the rotating JNW spacetime can have positive as well as negative (or opposite) precession of the timelike bound orbits.  
\end{itemize}  
In the context of the observations by the EHT, GRAVITY, SINFONI, and UCLA galactic center groups, these studies on shadows and precession of timelike bound orbits are very useful. These novel features of the shadow shape (prolate arc) and precession of timelike bound orbits (negative precession) in rotating JNW spacetime could be observationally significant to differentiate the naked singularities from the black holes.
\begin{widetext}
\section{Appendix}
\label{Appendix}
We discuss about the energy conditions of the metric (\ref{Rotating_JNW_NJA}) obtained using the original NJA. Consider the inverse form of the spacetime metric.

\begin{eqnarray}
    \partial_s^2 = -\left(1-\frac{2Mr}{\nu\rho^2} \right)^{-\nu} \left\{1 - \frac{a^2\sin^2\theta}{\Delta} \left[1 - \left(1-\frac{2Mr}{\nu\rho^2} \right)^{\nu} \right]^2 \right\} \partial_t^2 + \frac{1}{\rho^2} \left(1-\frac{2Mr}{\nu\rho^2} \right)^{\nu-1} \big\{\Delta\partial_r^2 + \partial_\theta^2 \big\} \nonumber \\ -\, \frac{2a}{\Delta} \left[1 - \left(1-\frac{2Mr}{\nu\rho^2} \right)^{\nu} \right] \partial_t \partial_\phi + \frac{1}{\Delta\sin^2\theta} \left(1-\frac{2Mr}{\nu\rho^2} \right)^{\nu} \partial_\phi^2
\end{eqnarray}
It can be written as
\begin{eqnarray}
    \partial_s^2 = -\left(1-\frac{2Mr}{\nu\rho^2} \right)^{-\nu} \partial_t^2 + \frac{1}{\rho^2} \left(1-\frac{2Mr}{\nu\rho^2} \right)^{\nu-1} \big[\Delta\partial_r^2 + \partial_\theta^2 \big] + \left(1-\frac{2Mr}{\nu\rho^2} \right)^{\nu} \bigg[\frac{a\sin\theta}{\sqrt{\Delta}} \left\{1-\left(1-\frac{2Mr}{\nu\rho^2} \right)^{-\nu} \right\} \partial_t \nonumber \\ +\, \frac{1}{\sqrt{\Delta}\sin\theta} \partial_\phi \bigg]^2 \,\,\,
    \label{Inverse JNW NJA}
\end{eqnarray}
Now, we choose a set of orthonormal basis $\{u^\mu, e_r^\mu, e_\theta^\mu, e_\phi^\mu \}$, which satisfies $u^\mu u_\mu = -1$, $e_{i}^{\mu} (e_i)_\mu = 1$, $u^\mu (e_i)_\mu = 0$; where, $(i \to r,\theta,\phi)$.
\begin{eqnarray}
    && u^\mu = \left(\left(1-\frac{2Mr}{\nu\rho^2} \right)^{-\nu/2}, 0, 0, 0 \right) \\
    && e_r^\mu = \left(0, \frac{\sqrt{\Delta}}{\rho}\left(1-\frac{2Mr}{\nu\rho^2} \right)^{\frac{\nu-1}{2}}, 0, 0 \right)\\
    && e_\theta^\mu = \left(0, 0, \frac{1}{\rho}\left(1-\frac{2Mr}{\nu\rho^2} \right)^{\frac{\nu-1}{2}}, 0 \right) \\
    && e_\phi^\mu = \left(1-\frac{2Mr}{\nu\rho^2} \right)^{\nu/2} \left(\frac{a\sin\theta}{\sqrt{\Delta}} \left\{1-\left(1-\frac{2Mr}{\nu\rho^2} \right)^{-\nu} \right\}, 0, 0,\frac{1}{\sqrt{\Delta}\sin\theta} \right)
\end{eqnarray}
Using eq. (\ref{rho general}), (\ref{Pr general}), (\ref{Ptheta general}), (\ref{Pphi general}), we determine the components of the energy momentum tensor. For mathematical simplicity we take $\theta=\pi/2$.
\begin{eqnarray}
    && \rho_e = P_r + \frac{2a^2M^2(1-\nu)^2}{r^6\nu^2} \left(1-\frac{2M}{r\nu} \right)^{-3+\nu} \\
    && P_r = - P_\theta = \frac{M^2 (1-\nu^2)}{r^4 \nu^2} \left(1 - \frac{2M}{r\nu} \right)^{-2+\nu} \label{Pr Ptheta}\\
    && P_\phi = P_\theta - \frac{2a^2M^2(1-\nu^2)}{r^6\nu^2} \left(1-\frac{2M}{r\nu} \right)^{-3+\nu}
\end{eqnarray}
We see that
\begin{equation}
    \rho_e + P_r + P_\theta + P_\phi = \rho_e + P_\phi = -\frac{4a^2M^2(1-\nu)}{r^6\nu} \left(1-\frac{2M}{r\nu} \right)^{-3+\nu} < 0
\end{equation}
which implies that the spacetime metric (\ref{Rotating_JNW_NJA}) does not obey the weak energy condition, null energy condition, and the strong energy condition. We note that the dominant energy condition is also violated. Thus, the rotating spacetime metric obtained using the Newman-Janis algorithm with complexification is not physically a valid solution of the Einstein field equation.
\end{widetext}

\end{document}